%% file: manuscript.tex
\documentclass[twocolumn]{aastex63}
\usepackage{booktabs}
\usepackage{longtable}
\usepackage{xspace}
\usepackage[T1]{fontenc}

\newcommand{\Kepler}{\textit{Kepler}\xspace} 
\newcommand{\spitzer}{\textit{Spitzer}\xspace} 
\newcommand{\ktwo}{\textit{K2}\xspace}

\newcommand{\Mstar}{\ensuremath{M_{\star}}\xspace}
\newcommand{\Rstar}{\ensuremath{R_{\star}}\xspace} 
 
\newcommand{\fe}{[Fe/H]\xspace}
\newcommand{\teff}{\ensuremath{T_{\mathrm{eff}}}\xspace}  
\newcommand{\logg}{\ensuremath{\log g}\xspace}

\newcommand{\Mp}{\ensuremath{M_p}\xspace} 
\newcommand{\Rp}{\ensuremath{R_p}\xspace}
\newcommand{\teq}{$T_{\mathrm{eq}}$\xspace}
\newcommand{\Sinc}{\ensuremath{S_{inc}}\xspace}
\newcommand{\fenv}{\ensuremath{f_{\mathrm{env}}}\xspace}
\newcommand{\menv}{\ensuremath{M_{\mathrm{env}}}\xspace}
\newcommand{\mcore}{\ensuremath{M_{\mathrm{core}}}\xspace}

\newcommand{\ms}{m s$^{-1}$\xspace}

\newcommand{\msyr}{m s$^{-1}$ yr$^{-1}$\xspace}
\newcommand{\Se}{\ensuremath{S_{\oplus}}\xspace}
\newcommand{\Me}{\ensuremath{M_{\oplus}}\xspace} 
\renewcommand{\Re}{\ensuremath{R_{\oplus}}\xspace} 
\newcommand{\um}{\ensuremath{\mu \mathrm{m}}\xspace}

\newcommand{\Rsun}{\ensuremath{R_{\odot}}\xspace }
\newcommand{\Msun}{\ensuremath{M_{\odot}}\xspace}

\def\deg{\ensuremath{^{\circ}}}

\newcommand{\sqrtecosw}{\ensuremath{\sqrt{e} \cos \omega}\xspace}
\newcommand{\sqrtesinw}{\ensuremath{\sqrt{e} \sin \omega}\xspace}
\newcommand{\dvdt}{\ensuremath{dv/dt}\xspace}
\newcommand{\sigjit}{\ensuremath{\sigma_\mathrm{jit}}\xspace}

\newcommand{\K}[1]{\ensuremath{K_\mathrm{#1}}\xspace}
\newcommand{\e}[1]{\ensuremath{e_\mathrm{#1}}\xspace}

\usepackage{xstring} % Needed for IfEqCase
\usepackage{ifthen} % needed for ifthenelse command
\usepackage{xifthen} % needed for ifthenelse command

\newcommand{\argperi}[1]{\ensuremath{\ifthenelse{\isempty{#1}}{\omega_P}{\omega_{P,#1}}}\xspace}
\newcommand{\inc}[1]{\ensuremath{\ifthenelse{\isempty{#1}}{i}{i_{#1}}}\xspace}
\newcommand{\ecc}[1]{\ensuremath{\ifthenelse{\isempty{#1}}{e}{e_{#1}}}\xspace}
\newcommand{\per}[1]{\ensuremath{\ifthenelse{\isempty{#1}}{P}{P_{#1}}}\xspace}
\newcommand{\node}[1]{\ensuremath{\ifthenelse{\isempty{#1}}{\Omega}{\Omega_{#1}}}\xspace}
\newcommand{\meananom}[1]{\ensuremath{\ifthenelse{\isempty{#1}}{M}{M_{#1}}}\xspace}
\newcommand{\ecosw}[1]{\ensuremath{\ifthenelse{\isempty{#1}}{e \cos \omega_P}{e \cos \omega_{#1}}}\xspace}
\newcommand{\esinw}[1]{\ensuremath{\ifthenelse{\isempty{#1}}{e \sin \omega_P}{e \sin \omega_{#1}}}\xspace}
\newcommand{\T}[1]{\ensuremath{\ifthenelse{\isempty{#1}}{T}{T_{#1}}}\xspace}
\newcommand{\tc}[1]{\ensuremath{\ifthenelse{\isempty{#1}}{T_{c}}{T_{c,#1}}}\xspace}
\newcommand{\secosw}[1]{\ensuremath{\sqrt{e_{#1}} \cos \omega_{#1}}\xspace}
\newcommand{\sesinw}[1]{\ensuremath{\sqrt{e_{#1}} \sin \omega_{#1}}\xspace}
\newcommand{\rrat}[1]{\ensuremath{R_{p,#1}/\Rstar}\xspace}

\newcommand{\zfree}[1]{\ensuremath{\ifthenelse{\isempty{#1}}{z_{\mathrm{free}}^{*}}{z_{\mathrm{free,#1}}}}\xspace}

\newcommand{\dbic}{\ensuremath{\Delta\mathrm{BIC}}\xspace}

\input{val_stat.tex}
\input{val_sys.tex}

\begin{document}

\title{K2-19b and c are in a 3:2 Commensurability but out of Resonance: \\ A Challenge to Planet Assembly by Convergent Migration}

\author{Erik~A.~Petigura}
\affiliation{Department of Physics \& Astronomy, University of California Los Angeles, Los Angeles, CA 90095, USA}

\author{John~Livingston}
\altaffiliation{JSPS Fellow}
\affiliation{Department of Astronomy, University of Tokyo, 7-3-1 Hongo, Bunkyo-ku, Tokyo 113-0033, Japan}

\author{Konstantin~Batygin}
\affiliation{Division of Geological and Planetary Sciences, California Institute of Technology, Pasadena CA, 91125, USA}

\author{Sean~M.~Mills}
\affiliation{Cahill Center for Astrophysics, California Institute of Technology, Pasadena CA, 91125, USA}

\author{Michael~Werner} 
\affiliation{Jet Propulsion Laboratory, California Institute of Technology, Pasadena, CA 91109, USA}

\author{Howard~Isaacson}
\affiliation{Department of Astronomy,  University of California Berkeley, Berkeley CA 94720}
\affiliation{University of Southern Queensland, Toowoomba, QLD 4350, Australia}

\author{Benjamin~J.~Fulton} 
\affiliation{IPAC-NASA Exoplanet Science Institute Pasadena, CA 91125, USA}

\author{Andrew~W.~Howard}
\affiliation{Cahill Center for Astrophysics, California Institute of Technology, Pasadena CA, 91125, USA}

\author{Lauren~M.~Weiss} 
\altaffiliation{Parrent Postdoctoral Fellow}
\affiliation{Institute for Astronomy, University of Hawaii at Manoa, Honolulu, HI 96822, USA}

\author{N\'estor~Espinoza} 
\altaffiliation{Bernoulli Fellow}
\altaffiliation{IAU-Gruber Fellow}
\affiliation{Max-Planck-Institut f\"ur Astronomie, K\"onigstuhl 17, 69117 Heidelberg, Germany}

\author{Daniel~Jontof-Hutter} 
\affiliation{Department of Physics, University of the Pacific, Stockton, CA 95211, USA}

\author{Avi~Shporer} 
\affiliation{Department of Physics and Kavli Institute for Astrophysics and Space Research, Massachusetts Institute of Technology, Cambridge, MA 02139, USA}

\author{Daniel~Bayliss} 
\altaffiliation{Observatoire Astronomique de l'Universit\'e de Gen\`eve, 51 ch. des Maillettes, 1290 Versoix, Switzerland}

\author{S.~C.~C.~Barros}
\affiliation{Instituto de Astrof\'isica e Ci\^encias do Espa\c{c}o, Universidade do Porto, CAUP, Rua das Estrelas, PT4150-762 Porto, Portugal}

\correspondingauthor{Erik A. Petigura}
\email{petigura@astro.ucla.edu}

\begin{abstract}
\object{K2-19}b and c were among the first planets discovered by NASA's \ktwo mission and together stand in stark contrast with the physical and orbital properties of the solar system planets. The planets are between the size of Uranus and Saturn at \sys{pd-prad2_fmt}~\Re and \sys{pd-prad3_fmt}~\Re, respectively, and reside a mere 0.1\% outside the nominal 3:2 mean-motion resonance. They represent a different outcome of the planet formation process than the solar system, as well as the vast majority of known exoplanets. We measured the physical and orbital properties of these planets using photometry from \ktwo, \spitzer, and ground-based telescopes, along with radial velocities from Keck/HIRES. Through a joint photodynamical model, we found that the planets have moderate eccentricities of $e \approx0.20$ and well-aligned apsides $\Delta \varpi \approx 0$~deg. The planets occupy a strictly non-resonant configuration: the resonant angles circulate rather than librate. This defies the predictions of standard formation pathways that invoke convergent or divergent migration, both of which predict $\Delta \varpi \approx 180$~deg and eccentricities of a few percent or less. We measured masses of $M_{p,b}$ = \sys{pd-masseprec2_fmt}~\Me and $M_{p,c}$ = \sys{pd-masse3_fmt}~\Me. Our measurements, with $5\%$ fractional uncertainties, are among the most precise of any sub-Jovian exoplanet. Mass and size reflect a planet's core/envelope structure. Despite having a relatively massive core of $\mcore \approx15$~\Me, K2-19b is envelope-rich, with an envelope mass fraction of roughly 50\%. This planet poses a challenge to standard models core-nucleated accretion, which predict that cores $\gtrsim 10$~\Me will quickly accrete gas and trigger runaway accretion when the envelope mass exceeds that of the core.
\end{abstract}


\section{Introduction}
While a perennial quest in exoplanet astronomy is the discovery and characterization of ever more ``Earth-like'' worlds, our understanding of planet formation is best informed by the full diversity of planets around other stars. Thanks to the rapidly growing census of extrasolar planets, we may now study the diverse outcomes of planet formation  processes beyond those that occurred in the solar system. The K2-19 system is one such outcome.

The system hosts three known planets. \cite{Armstrong15} initially reported K2-19b and c based on photometry collected by the {\em Kepler Space Telescope} operating in its \ktwo mode \citep{Howell14}. K2-19b has an orbital period of 7.9~days and has a radius of \sys{pd-prad2}~\Re, between the size of Uranus and Saturn. K2-19c has an orbital period of 11.9~days and a radius of \sys{pd-prad3}~\Re. While K2-19c is similar in size to the solar system ice giants, aspects of its bulk composition, such as ice fraction, may be quite different due its close-in orbit. As techniques to correct for \ktwo systematics improved, \cite{Sinukoff16} detected a third planet, K2-19d, a 1.2~\Re planet on 2.5 day orbit.

In this paper, we focus on K2-19b and c, which reside just outside the nominal 3:2 mean-motion resonance. While \cite{Armstrong15} detected transit-timing variations (TTVs) within the \ktwo dataset, the relatively short 80~day baseline resulted in significant uncertainties in the TTV model. Several groups have subsequently observed transits of K2-19b from the ground in order to better constrain the TTV model \citep{Armstrong15,Narita15,Barros15}. However, to date, there have been no successful recoveries of the K2-19c transit, which has contributed to lingering uncertainty in the TTV solution.

In parallel, several groups have obtained radial velocity (RV) measurements of K2-19 in order to directly constrain the planet masses through stellar reflex motion \citep{Dai16,Nespral17}. A key challenge to these efforts is that at $V = 13.0$~mag, K2-19 is near the faint limit of most current RV facilities. In addition, the star exhibits significant RV variability due to spot modulation, which must be disentangled from the planetary signals.

In this work, we present the results of a coordinated observational campaign to characterize K2-19b and c, using both TTVs and RVs. We describe our photometry in Section~\ref{sec:photometry} and our RVs in Section~\ref{sec:rv}. Our photometric dataset includes two \spitzer observations for each planet.  Our \spitzer observations of the K2-19c transits are significant in that they are the first \ktwo and help to reduce uncertainties in the TTV solution. We perform a photodynamical analysis in Section~\ref{sec:analysis}, which yields the most precise constraints on the masses and orbits of these two planets to date. In Sections~\ref{sec:core-envelope}--\ref{sec:formation} we assess the bulk composition of these planets, their dynamical evolution, and possible formation pathways.

\section{Photometric Observations}
\label{sec:photometry}

\subsection{K2 Photometry}
\label{sec:obs-ktwo}
\begin{figure*}[h!]
\centering
\includegraphics[width=0.97\textwidth]{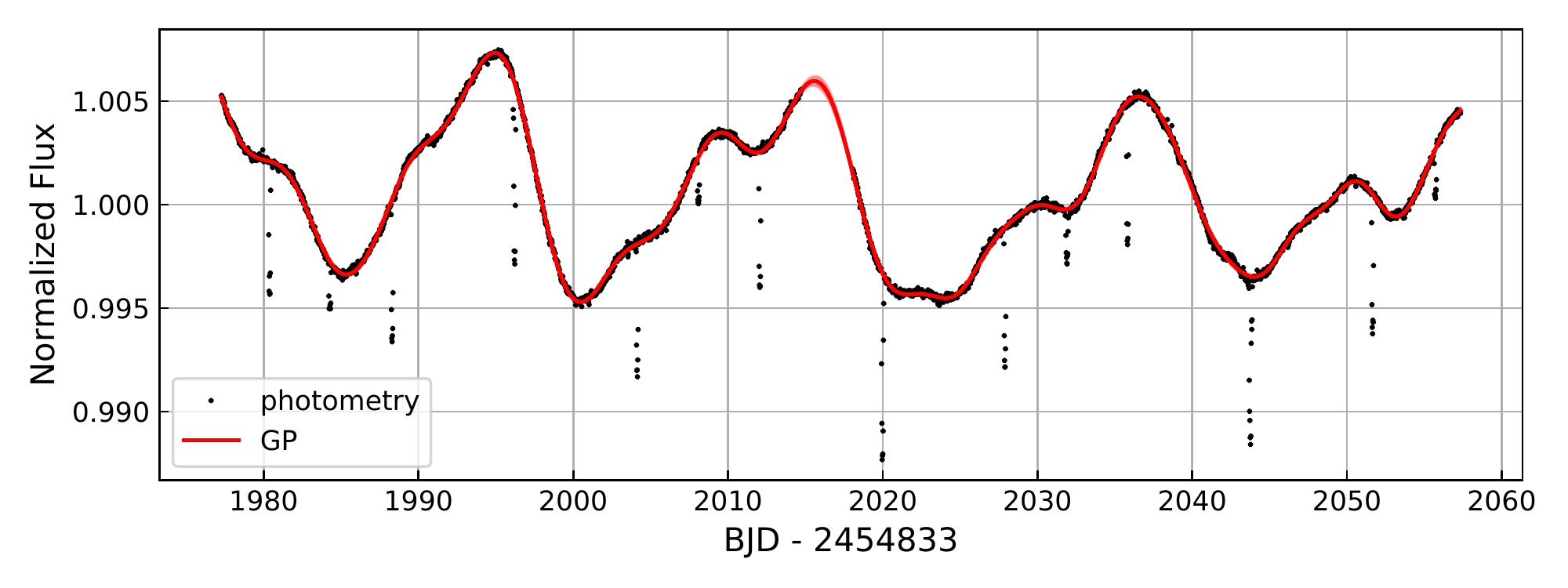}
\caption{Photometry from \ktwo after removing instrument systematics  showing $\approx1\%$ periodic variability with $P \approx20$~days (see Section~\ref{sec:obs-ktwo}). The red line is our Gaussian Process fit to the photometry, which informs the adopted noise model in our RV analysis (see Section~\ref{sec:rv-keplerian}).\label{fig:lightcurve-ktwo}}
\end{figure*}

The {\em Kepler Space Telescope} observed K2-19 from 2014-05-30 to 2014-8-21 during campaign 1 of its \ktwo mission. The photometry contain large systematics due to pointing drifts of $\sim$1~pixel that occur on $\sim$6~hr timescales. We used the EVEREST2.0 package to correct for these systematics \citep{Luger17}, and the corrected light curve is shown in Figure~\ref{fig:lightcurve-ktwo}.%

\begin{figure*}
\centering
\includegraphics[width=0.97\textwidth]{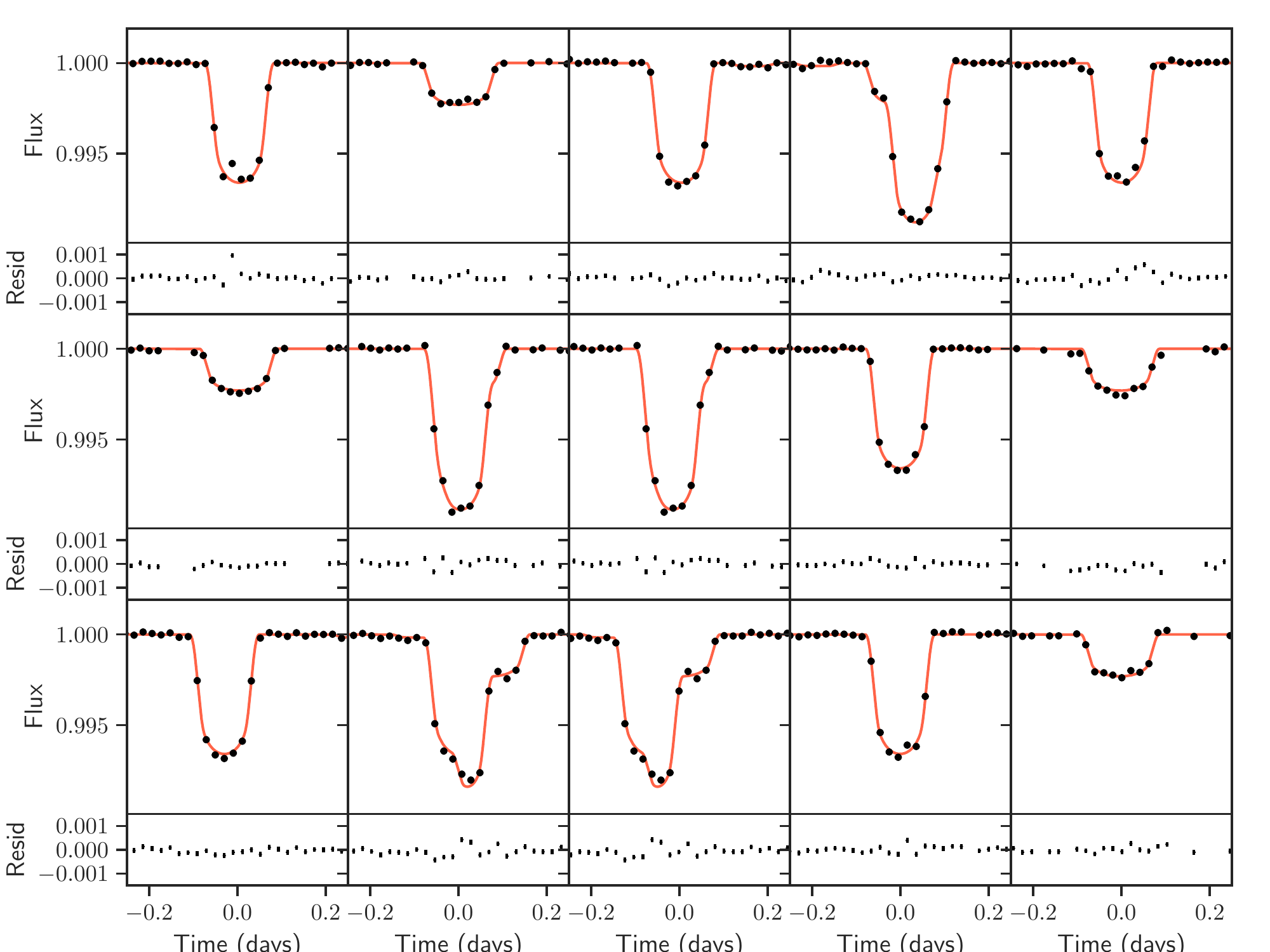}
\caption{The black circles show the detrended \ktwo photometry around the transits. Several overlapping transits are observed. The maximum a posteriori model is shown as the orange line and the residuals to this model are shown below. Increased scatter during transit due to spot crossing events are observed during some transits (see Section~\ref{sec:photodyn}).
\label{fig:photodyn-fit}}
\end{figure*}

There is clear periodic variability with $P \approx 20$~days with a peak-to-trough amplitude of 1\% due to rotation-induced spot modulation. Figure~\ref{fig:photodyn-fit} is a zoomed in view of individual transits, some of which are overlapping.

\subsection{Spitzer Photometry}
\label{sec:obs-spitzer}
The \ktwo data alone samples only a small fraction of the multi-year TTV signal. We used the {\em Spitzer Space Telescope} to observe two additional transits of K2-19b and K2-19c to better sample this  signal. Planet b observations were conducted on 2017-04-23 and 2017-09-05; planet c observations were conducted on 2016-10-04 and 2017-04-08.%
\footnote{All observations were carried out under GO program 13052 (PI: M. Werner).}
%.

To plan the first set of \spitzer observations, we consulted the transit times predicted by \cite{Barros15} (S.~Barros, private communication). Because our first \spitzer observation of K2-19b was two years after the last transit used in the \cite{Barros15} model, there was considerable timing uncertainty. We observed for 12 hours to reliably catch the 3.5~hour transit.  There was even more timing uncertainty for K2-19c, which had not been observed since 2014, and we scheduled a 27~hour observing sequence.

When planning our second set of \spitzer observations, we constructed a preliminary TTV model with plausible values for the planet masses and eccentricity. Having incorporated the first set of observations, there was less uncertainty in the transit times of K2-19b and c, requiring only 7 and 9 hour observing sequences, respectively.

We used IRAC channel 2 (4.5~$\mu$m) because the instrumental systematics due to intra-pixel sensitivity variations are smaller than in channel 1 (3.6~$\mu$m; \citealt{Ingalls12}). We used 2 second exposures to optimize the integration efficiency while remaining in the linear regime of the detector. We extracted photometry from the \spitzer data using circular apertures. As described in \cite{Livingston19}, we selected the aperture size ($r$ = 2.2~pixels) that minimized the combined uncorrelated (white) and correlated (red) noise, as measured by the standard deviation and $\beta$ factor \citep{Pont06,Winn08}. We resampled the light curve into 60 second integrations which yields improved systematic modeling without significantly altering the transit profile \citep{Benneke17}.

Following standard practice, we modeled the \spitzer systematics and transit profile simultaneously. Using the pixel-level decorrelation (PLD) method of \cite{Deming15}, we constructed our systematic model from a linear combination of the nine pixel-level lightcurves from a $3 \times 3$ pixel grid centered on the star. For K2-19b and c, we modeled each set of two transits simultaneously and shared all transit parameters except for the transit mid-times $\tc{i}$. We used a quadratic limb-darkening parameterization and physically motivated priors \citep{Claret12,Kipping13}. In summary, our modeling of each planet involved 28 free parameters: nine PLD coefficients for each dataset, a white noise term for each dataset, two transit mid-points $\tc{i}$, the orbital period $P$, the planet-star radius ratio $\Rp/R_{\star}$, the scaled semi-major axis $a/R_{\star}$, the impact parameter $b$, the limb-darkening parameters $q_1$ and $q_2$. 

\begin{figure*}
\centering
\includegraphics[trim={0cm 0cm 0cm 0cm},width=0.9\textwidth]{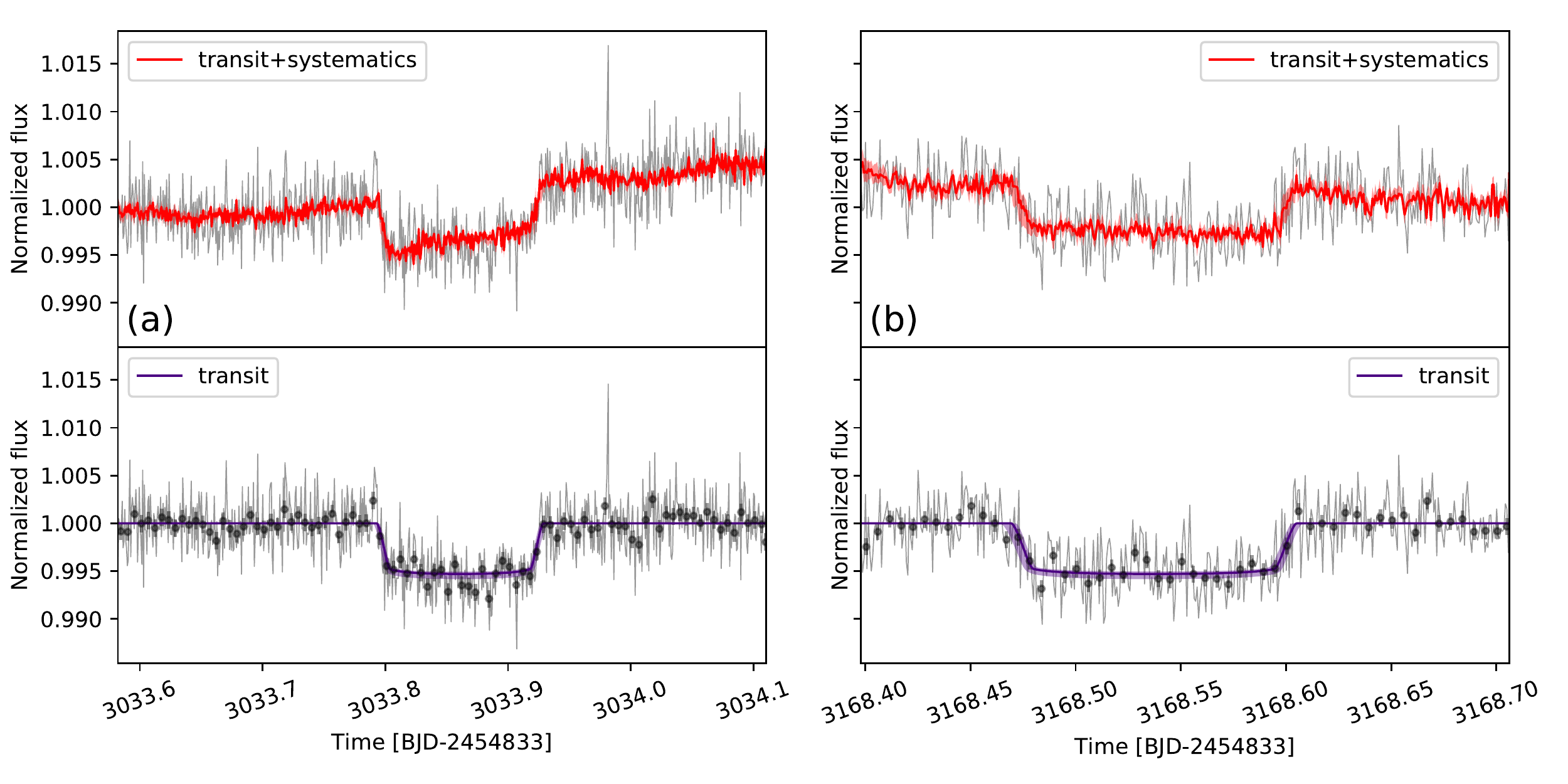}
\\[-1ex]
\includegraphics[trim={0cm 0cm 0cm 0cm},width=0.9\textwidth]{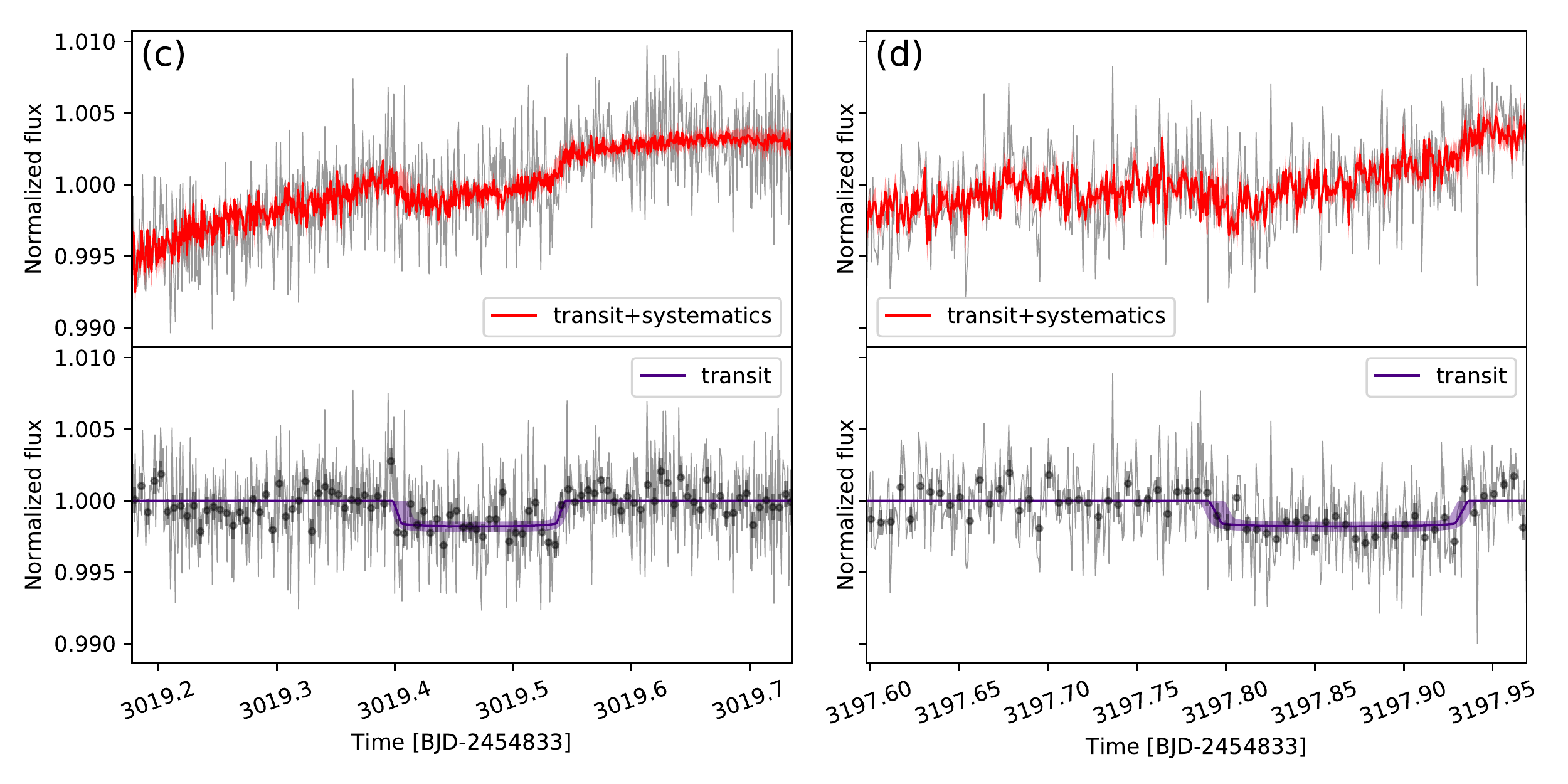}
\caption{Transits of K2-19b and c observed by \spitzer in the 4.5~\um IRAC channel along with our simultaneous modeling of the instrumental systematics and transit profiles (see Section~\ref{sec:obs-spitzer}).  Panel (a) shows the transit of K2-19b with a transit number $i=133$, where $i=0$ corresponds to the first \ktwo transit. In the top sub-panel, we show the raw light curve (gray), the maximum a posteriori (MAP) transit/systematic model (red), and the 95\% credible models (light red band). In the bottom sub-panel, we show the MAP corrected photometry (gray) and transit model (purple). The 95\% credible models are shown with the purple band. Panel (b): same as (a) but for the second \spitzer observation of K2-19b ($i=150$). Panel (c): same as (a) but for the first \spitzer observation of K2-19c ($i=87$). Panel (d): same as (a) but for the second \spitzer observation of K2-19c ($i=102$).\label{fig:lightcurve-spitzer}}
\end{figure*}

We explored the range of coefficients allowed by our data using the affine-invariant Markov Chain Monte Carlo (MCMC) sampler of \cite{Goodman10}. We initialized 100 walkers and allowed them to evolve for $N_\mathrm{steps} = 10000$ steps. We visually inspected the trace plots and discarded the first 5000 steps of burn-in. We assessed the convergence by computing the autocorrelation length $\tau$ for each chain. We computed the mean value of $\tau$ for all 100 chains for each parameter, and found that $N_\mathrm{steps}/\tau \geq 38$ for all chains. Our corrected light curves had an RMS scatter of $\sim$400 ppm on 40 minute timescales. The \spitzer photometry and best-fit transit models are shown Figure~\ref{fig:lightcurve-spitzer}. The derived transit times are listed in Table~\ref{tab:transit-times}.

\begin{deluxetable}{lllrrrr}
\tablecaption{Transit Times\label{tab:transit-times}}
\tabletypesize{\footnotesize}
\tablehead{
  \colhead{Planet} & 
  \colhead{Transit} & 
  \colhead{Instrument} & 
  \colhead{$T_{c}$} & 
  \colhead{$\sigma(T_{c})$} &
  \colhead{Notes}   \\
  \colhead{} & 
  \colhead{} & 
  \colhead{} & 
  \colhead{days} & 
  \colhead{days} &
  \colhead{}
}
\startdata
\input{tab_times.tex}
\enddata
\tablecomments{Following a convention from the \Kepler mission, times are given in $\mathrm{BJD}_\mathrm{TBD} - 2454833$. Notes---A: This work; B: \cite{Narita15}}
\end{deluxetable}

\subsection{Ground-based Photometry}
\label{sec:obs-ground}

\begin{figure}[h]
\centering
\includegraphics[trim={0cm 0cm 0cm 0cm},width=0.49\textwidth]{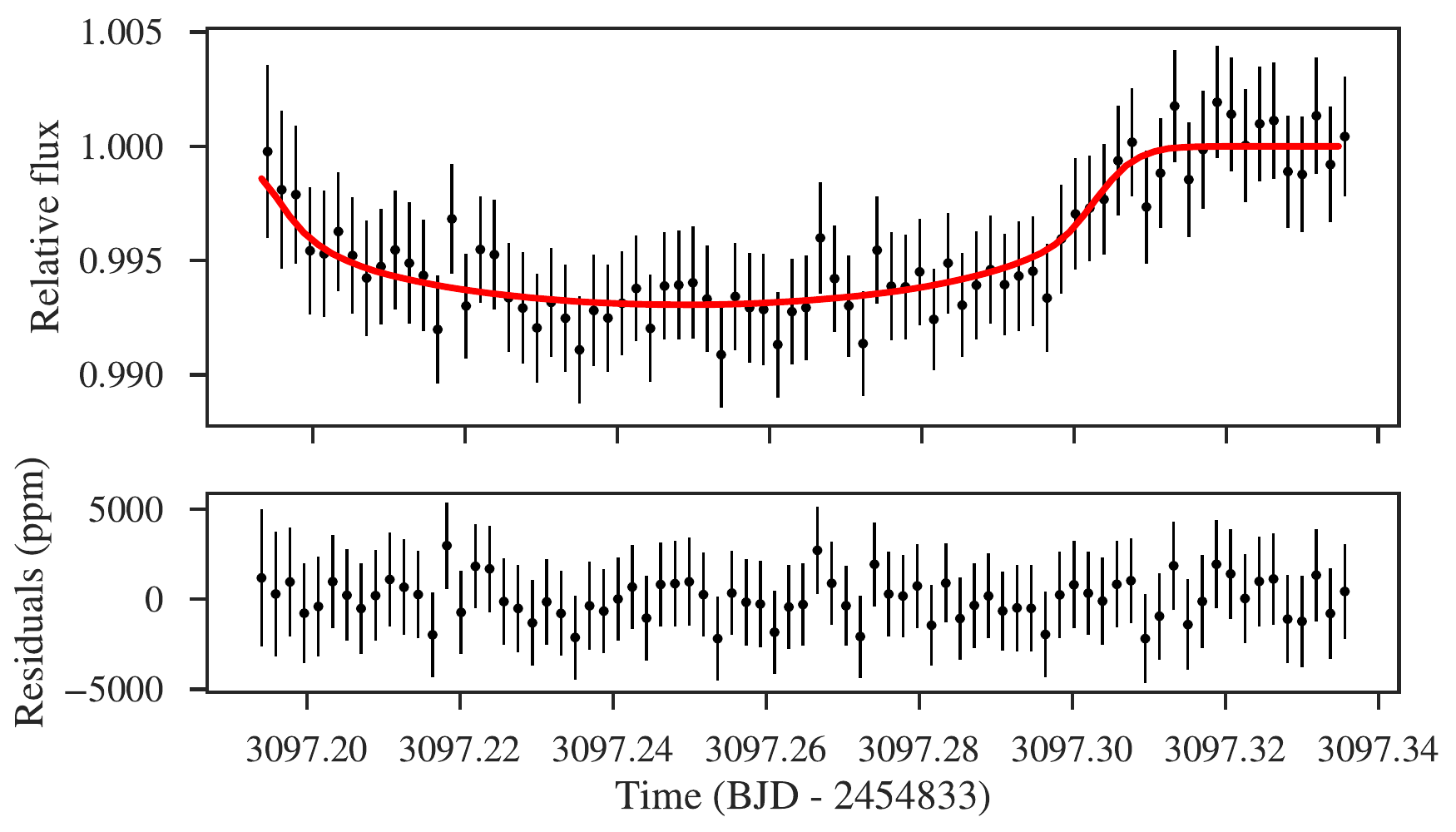}
\caption{Top panel: black points show the relative photometry of K2-19 observed by LCO on 2017-06-05 during the transit of K2-19b (see Section~\ref{sec:obs-ground}). The red line is the best fit transit model and the bottom panel shows the residuals to the fit.\label{fig:photometry-lco}}
\end{figure}

We also included several transit times of K2-19b measured using ground-based facilities. Three were drawn from \cite{Narita15}. We also observed a transit on 2017-06-05 with the 1m telescope of the Las Cumbres Observatory network (LCO; \citealt{Brown13}), located at the South African Astronomical Observatory. We performed bias, dark, and flat-field corrections using the standard LCOGT pipeline \citep{BANZAI}. We then performed aperture photometry on K2-19 and 10 comparison stars having similar 2MASS colors and performed differential photometry to remove instrumental and atmospheric effects. We modeled the transit using both white and correlated noise models and found that the white-noise mode was preferred. The light curve and transit fit are shown in Figure~\ref{fig:photometry-lco}. The RMS scatter in the residuals is $\approx 300$~ppm per 40 min interval.

% It's approximately 1000ppm per 3 min measurement or 273 ppm per 40 min

\section{Radial Velocity Observations}
\label{sec:rv}
We obtained \stat{rv-n} spectra of K2-19 using the High Resolution Echelle Spectrometer (HIRES; \citealt{Vogt94}) on the 10m Keck-I telescope between \stat{rv-start} and \stat{rv-stop}. We collected spectra through an iodine cell mounted directly in front of the spectrometer slit. The iodine cell imprints a dense forest of absorption lines which serve as a wavelength reference.  We also obtained a ``template'' spectrum without iodine.

At $V = 13.0$~mag K2-19 is a challenging RV target for Keck/HIRES. We aimed to achieve a consistent signal-to-noise ratio (SNR) of 100 per reduced pixel at 5500~\AA\ using an exposure meter. However, various throughput losses due to poor/variable seeing and cirrus clouds sometimes resulted in lower than desired SNR. Our spectra have per pixel SNR ranging from \stat{rv-snr-min} to \stat{rv-snr-max}.

RVs were determined using standard procedures of the California Planet Search \citep{Howard10b} including forward modeling of the stellar and iodine spectra convolved with the instrumental response (\citealt{Marcy92,Valenti95}). The measurement uncertainty of each RV point is derived from the uncertainty on the mean RV of the $\sim$700 spectral chunks used in the RV pipeline and ranges from \stat{rv-errvel-min} to \stat{rv-errvel-max} \ms. Table~\ref{tab:rv} lists the RVs and uncertainties.

\begin{deluxetable}{RRRRRr}
\tablecaption{Radial Velocities\label{tab:rv}}
\tablecolumns{5}
\tablewidth{-0pt}
\tabletypesize{\footnotesize}
\tablehead{
  \colhead{Time} & 
  \colhead{RV} & 
  \colhead{$\sigma$(RV)} &
  \colhead{$\mathrm{S}_\mathrm{HK}$} \\ 
  \colhead{days} & 
  \colhead{\ms} & 
  \colhead{\ms} & 
  \colhead{}  
}
\startdata
\input{tab_rv-stub.tex}
\enddata
\tablecomments{Radial velocities and uncertainties for K2-19 (see Section~\ref{sec:rv}). Times are given in $\mathrm{BJD}_\mathrm{TBD} - 2454833$.  We also provide the Mount Wilson $\mathrm{S}_\mathrm{HK}$ activity index \citep{Vaughan78}, which is measured to 1\% precision. Table \ref{tab:rv} is published in its entirety in machine-readable format. A portion is shown here for guidance regarding its form and content.}
\end{deluxetable}

\section{TTV and RV Modeling}
\label{sec:analysis}

Here, we describe our modeling of both the photometric and RV datasets. In Section~\ref{sec:rv-keplerian}, we perform a Keplerian analysis of the RVs only. We observe quasiperiodic RV variability due to rotating starspots, which we model with a Gaussian process. Section~\ref{sec:photodyn} describes our photodynamical analysis that incorporates constraints from both photometry and RVs. This analysis yields tighter constraints on the properties of K2-19b and c and the parameters listed in Table~\ref{tab:sys} constitute our adopted system parameters.

While the photodynamical analysis yields smaller uncertainties, we present the RV-only analysis for the following reasons: (1) The RVs provide sensitivity to non-transiting planets that could compromise the accuracy of the photodynamical model. (2) RV variability from rotating starspots is comparable in amplitude to that due to K2-19b and may account for discrepancies between previously published mass measurements. (3) The two analyses demonstrate the relative strengths and weaknesses of the TTV and RV techniques as probes of the properties of the K2-19 system.

\subsection{Keplerian RV modeling}
\label{sec:rv-keplerian}
We analyzed the RV timeseries using the open source package RadVel \citep{Fulton18a}. RadVel facilitates maximum a posteriori (MAP) model fitting and parameter estimation via MCMC. In general, a Keplerian RV signal may be described by the orbital period $P$, time of inferior conjunction \tc{}, eccentricity $e$, argument of periastron $\omega$, and Doppler semi-amplitude $K$. We included K2-19b, c, and d in our model with zero eccentricity. For planets b and c we fixed $P$ and $\tc{}$ to the mean value as determined by the \ktwo and \spitzer photometry. For planet d, we fixed $P$ and $\tc{}$ to the \cite{Sinukoff16} ephemeris. While the planets do not have strictly linear ephemerides, we confirmed that the errors introduced by this simplification are negligible after performing the photodynamical analysis described in Section~\ref{sec:photodyn}. In our preliminary fitting, we found that models with a linear acceleration term \dvdt were favored by the Bayesian Information Criterion (BIC; \citealt{Schwartz78}) with $\dbic = -15$. In our subsequent modeling, described below, we found \dvdt = \sys{rv-dvdt_fmt}~\msyr.

The K2-19 photometry shows clear spot modulation (see Figure~\ref{fig:lightcurve-ktwo}), which can introduce correlated noise into the RV timeseries. We estimated the amplitude of this noise using the $F F^{\prime}$ method of \cite{Aigrain12}:
\begin{eqnarray}
\nonumber
\Delta \mathrm{RV} & \sim & F F^{\prime} \Rstar / f \\
\nonumber
                   & \sim & (0.5\%) (1\% / 5\, \mathrm{d}) (0.82\, \Rsun) / (1\%) \\
\nonumber
                   & \sim & 7 \, \mathrm{m s}^{-1}.\nonumber
\end{eqnarray}
Here, $F$ is the fractional flux variation, $F^{\prime}$ is its time derivative, and $f$ is the maximum flux decrement due to spots. This noise source is quasiperiodic as spots rotate with the stellar photosphere and also evolve with time. Numerous prior studies have modeled spot noise with quasiperiodic Gaussian Processes (GPs) including \cite{Haywood14}, \cite{Grunblatt15}, and others. We used the following quasiperiodic kernel that specifies the covariance between the $i$ and $j$ measurements:
\begin{eqnarray}
\nonumber
C_{i,j}  &=& \eta_1^2
\mathrm{exp}
	\left[ 
		-\frac{(t_i - t_j)^2}{\eta_2^2} 
		-\frac{1}{2\eta_4^2} \sin^2 \frac{\pi (t_i - t_j)}{\eta_3^2}
	\right] \\
\nonumber
   &+&
	\left[ 
		\sigma_i^2 + \sigjit^2
	\right]  
\delta_{i,j}.
\end{eqnarray}
Here, $\eta_1$ is the covariance amplitude, $\eta_2$ is the exponential decay length, $\eta_3$ sets the period, $\eta_4$ sets the relative importance of the exponential decay part of the kernel, and $\delta_{i,j}$ is the Kronecker delta function. We trained the GP on the \ktwo photometry and found $\eta_1 = \sys{phot-gp-eta1} \pm \sys{phot-gp-eta1_err}\%$, $\eta_2 = \sys{phot-gp-eta2} \pm \sys{phot-gp-eta2_err}$~days, $\eta_3 = \sys{phot-gp-eta3} \pm \sys{phot-gp-eta3_err}$~days, and $\eta_4 = \sys{phot-gp-eta4} \pm \sys{phot-gp-eta4_err}$. Our value for $\eta_3$ is consistent with our visual assessment of the stellar rotation period of $P \approx 20$~days.

We then modeled the RVs using the GP-based likelihood (see RadVel documentation for details). We imposed Gaussian priors on $\eta_2$, $\eta_3$, and $\eta_4$ based on our photometric modeling described above.  In summary, our RV model had the following free parameters: $\{K_1, K_2, K_3, \dvdt, \eta_1, \eta_2, \eta_3, \eta_4, \gamma, \sigjit\}$. 

Figure \ref{fig:rv-keplerian} shows the MAP model. We derived uncertainties using MCMC, terminating the chains when the inter-ensemble GR statistic was less than 1.003. For K2-19b, we measured a mass of \sys{rv-mpsini1_fmt}~\Me. The RVs were insufficient to detect planaets c or d, but we placed upper limits on their masses of $M_{p,c}$~<~\sys{rv-mpsini2_p95}~\Me and $M_{p,d}$~<~\sys{rv-mpsini3_p95}~\Me at 95\% confidence. 

We found that $\eta_1$, the amplitude of the quasiperiodic RV variability included in our GP noise model was \sys{rv-gp_amp_fmt} \ms, in agreement with our previous estimate. This value is comparable to reflex velocity of planet b, and it underscores the importance of treating spot-induced RV-variability in the RV analysis. We recommend that future RV campaigns targeting K2-19 (or similar stars) observe at high cadence to better trace this quasiperiodic noise source.

We explored fits where $e_b$ and $\omega_b$ were allowed to vary. However, this additional model complexity was disfavored by the BIC, with $\Delta$BIC = $-5$. Therefore, the RVs alone are insufficient to detect eccentricity for K2-19b. We characterized the values of $e_b$ excluded solely by the RVs by running a second MCMC where \secosw{b} and \sesinw{b} were allowed to vary. We found that $e_b < 0.27$ at 95\% confidence, which is consistent with our photodynamical analysis presented in Section~\ref{sec:photodyn}.

We note that our RV-only mass measurement of planet b is inconsistent at the $\sim$$3\sigma$ level with that of \cite{Nespral17}, who reported $M_{p,b} = 54.8 \pm 7.5$~\Me. The \cite{Nespral17} analysis used 22 RVs from three different instruments: FIES, HARPS-N, and HARPS. We hypothesize that, in the \cite{Nespral17} analysis, biases due to stellar activity were amplified given the sparse sampling of the RV timeseries and offsets between the RV datasets. 

As we show in Section~\ref{sec:photodyn}, the constraints from TTVs on the masses and eccentricities of K2-19b and c are more precise than those from the RVs. However, the RVs provide sensitivity to non-transiting planets that could compromise the accuracy of the TTV model. Non-transiting planets near first order MMR are the most concerning, as they would produce the largest TTVs.

To search for such planets, we computed the Lomb-Scargle periodogram \citep{Lomb76,Scargle82} of the residuals to the most probable Keplerian model (see Figure~\ref{fig:rv-keplerian}). We found no additional signals with a bootstrap false alarm probability of < 10\% \citep{VanderPlas18}. Detection of an exoplanet from RVs alone with $P$ less than the observing baseline generally requires $K \gtrsim \alpha  \sigma_\mathrm{RV} / \sqrt{N_{obs}}$, where $\sigma_\mathrm{RV} $ is the individual RV measurement uncertainty and $\alpha$ is a numerical prefactor of $\approx$$10$ \citep{Howard16}. Adopting $\sigma_\mathrm{RV} = 9$~\ms, the quadrature sum of the two dominant noise terms, $\sigjit$ and $\eta_1$, we found that a planet with $K \gtrsim 13 $~\ms would have been detectable. Therefore, at orbital periods comparable to those of K2-19b and c, the RVs rule out planets with masses comparable to K2-19b. This supports the assumption in our photodynamical model that the TTV signal is dominated by interactions between K2-19b and K2-19c.

\begin{figure*}
\centering
\includegraphics[width=0.9\textwidth]{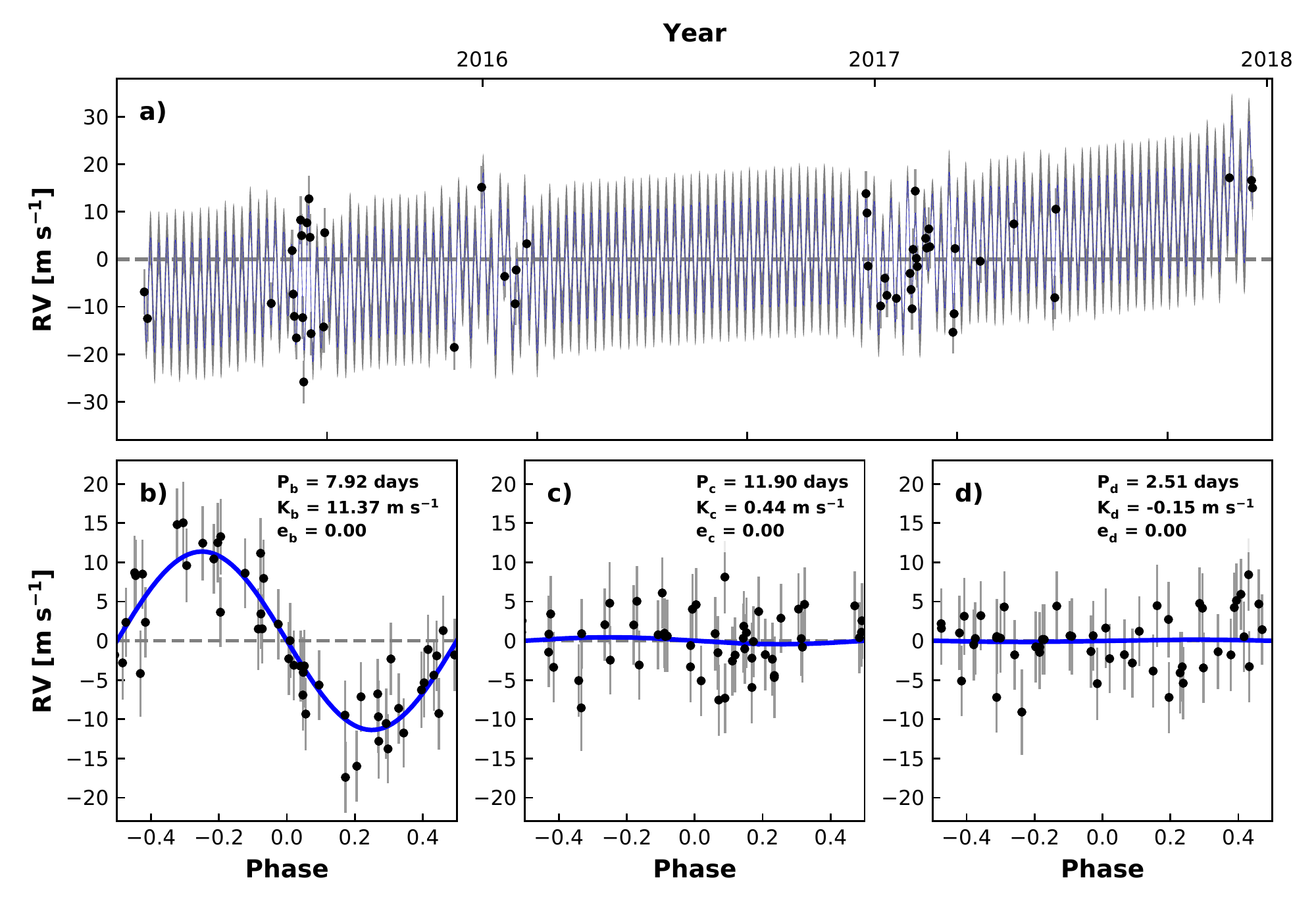}
\label{fig:rv-keplerian}
\caption{The three-Keplerian fit to the K2-19 radial velocities (RVs), assuming circular orbits (see Section~\ref{sec:rv-keplerian}). Panel (a): Points show RVs from HIRES and the line shows the most probable Keplerian model. The gray band shows Gaussian process model that accounts for quasiperiodic correlated noise due to star spots. Panel (b) shows the phase-folded RVs and the most probable Keplerian model for K2-19b with contributions from the GP noise model, $dv/dt$ term, and other Keplerians removed. Panel (c), same as (b), but for K2-19c. Panel (d), same as (b) but for K2-19d.\label{fig:rv-keplerian}}
\end{figure*}

\subsection{Photo-dynamical analysis}
\label{sec:photodyn}
To extract the information contained in both the RV and photometric datasets, we performed a photodynamical analysis. We used the Phodymm code, which is described in \cite{Mills16}. Given an initial configuration, Phodymm performs an $N$-body integration and forward models the light curve. The forward modeling approach has the advantage that it naturally handles simultaneous transits \citep{Pal08} and simultaneously models all transit characteristics such as duration and depth variations, compared to other techniques that model derived transit times (see, e.g., TTVFast; \citealt{Deck14}). 

For each planet, we specified an initial set of osculating elements: $P$, $\tc{}$, $e$, $\omega$, $i$, $\Omega$. Here, $i$ is the inclination and $\Omega$ is the longitude of ascending node. The model also requires \Mp and $\Rp/\Rstar$ for each planet, and the following stellar parameters: $\Mstar$, $\Rstar$, and quadratic limb-darkening parameters, $q_1$ and $q_2$. 

Because $\Omega$ is defined with respect to an arbitrary reference direction, we may fix $\Omega_b$ to 0 deg without loss of generality. K2-19d is dynamically decoupled from K2-19b and c and does not significantly affect the transits of the other planets gravitationally. However, K2-19d sometimes transits at the same time as K2-19b or c and therefore must be modeled out. We fixed $e_d$~=~0, $\omega_d$~=~0~deg, and $\Omega_d$~=~0~deg. Following the recommendations of \cite{Eastman13}, for planets b and c, we parameterized $\{e,\omega\}$ as $\{\sqrtecosw,\sqrtesinw\}$, which enforces a uniform prior on $e$. In total, our model had 24 free parameters.

% 8 * 3 + 4 - 3  (d fixed)   - 1 (c fixed) % 24

To assess the degree to which our model fits the \ktwo photometry, we defined
\[
\chi^2_{\mathrm{phot}} = \sum_{i} \left(\frac{f_{mod,i} - f_{i}}{\sigma_i}\right)^2, 
\]
where $f_{mod,i}$, $f_{i}$, $\sigma_i$ is the modeled flux, observed flux, and flux uncertainty of the $i^{\mathrm{th}}$ \ktwo observation.

For the \spitzer and ground based transits, we modeled the derived transit times (Table~\ref{tab:transit-times}) rather than the photometry directly because it is impractical to marginalize over the various systematic noise models that were used to derive the transit times. We defined the following goodness-of-fit statistic:
\[
\chi^2_{\mathrm{times}} = \sum_{j} \left(\frac{\tc{mod,j} - \tc{j}}{\sigma_j}\right)^2 
\]
where \tc{mod,j}, \tc{j}, and $\sigma_j$ are the modeled midpoint, observed midpoint, and timing uncertainty of the $j^{th}$ transit. Our final adopted log-likelihood is
\[
\log \mathcal{L} = -\frac{1}{2} \chi^2_{\mathrm{phot}} - \frac{1}{2} \chi^2_{\mathrm{times}}.
\]
Following \cite{Petigura18c}, we incorporated the RV mass constraints as Gaussian priors on the planet masses. We checked that this treatment is justified by verifying that the posteriors on $K_1$, $K_2$, and $K_3$ (Section~\ref{sec:rv-keplerian}) are Gaussian and uncorrelated. Finally, we applied Gaussian priors on $\Mstar$ and $\Rstar$ based on our stellar characterization (see Table~\ref{tab:sys}).

We explored the range of plausible models using  Differential Evolution Markov Chain Monte Carlo (DEMCMC). We ran 40 walkers and checked for convergence by periodically computing the Gelman-Rubin (GR) statistic \citep{Gelman92}. We terminated our runs after 80,000 steps, when the GR statistic was less than 1.05 for all parameters. After inspecting the chains, we discarded the first 10,000 steps as burn-in.

Figure~\ref{fig:photodyn-fit} shows the MAP photodynamical fit to the \ktwo dataset. We note that there is  increased scatter in the residuals during transits due to spot crossings events. These spot crossings do not systematically bias the model fits because they occur randomly over the transit chords. Figure~\ref{fig:photodyn-samples-times} shows 100 representative draws from the chains that illustrate the range of allowed transit times. The dominant TTV pattern is sinusoidal with $P^\prime \approx800$~days, but other harmonics are visible. To facilitate future observations of these planets we have included our predicted transit times 2029 in the Appendix.

The planet parameters are summarized in Table~\ref{tab:sys}. We have included a the joint posterior distributions for all parameters along with a discussion of several noteworthy covariances in the Appendix~\ref{sec:appendix}. We found that K2-19b and c are \sys{pd-masseprec2_fmt}~\Me and \sys{pd-masse3_fmt}~\Me, respectively.

While TTVs and RVs in principle provide complementary information, in our case, the TTVs are far more constraining. As an experiment, we ran the photodynamical model with no RV mass priors. The mass and eccentricity constraints are all consistent to within 2$\sigma$. In particular, photometry alone yields $M_{p,c} = 30.7\pm1.5$\Me. We note that K2-19c is one of roughly a dozen planets with independent mass constraints from TTVs and RVs. See \cite{Mills17} for further discussion and a comparison of the two techniques.

We show the constraints on the planets' eccentricity vectors ($e\cos \omega $, $e \sin \omega $) in Figure~\ref{fig:corner-ecc}. Both K2-19b and c have moderate eccentricities of $e_b$ =  \sys{pd-e2_fmt} and $e_c$ = \sys{pd-e3_fmt} and well-aligned apsides $\omega_c - \omega_b$ =  \sys{pd-omegadiffdeg_fmt}~deg. The eccentricities and orbital alignment of these two planets have important implications for formation history and their present-day dynamics, which we discuss in Section~\ref{sec:mmr}.

Previously, \cite{Barros15} measured masses and eccentricities of $M_{p,b}$ = $44 \pm 12$~\Me and $M_{p,c}$ = $16.9_{-2.8}^{+7.7}$~\Me and $e_b = 0.119^{+0.082}_{-0.035}$ and $e_c = 0.095^{+0.073}_{-0.035}$ using just the \ktwo photometry and three ground-based transits of K2-19b. Our measurements are consistent with those of \cite{Barros15} at the 1--2$\sigma$ level, but our measurements have smaller uncertainties on all parameters due to the additional \spitzer transits.

{\renewcommand{\arraystretch}{0.82}
\begin{table}
\begin{center}
\caption{K2-19 System Parameters}
\begin{tabular}{lrl}
\hline
\hline
Parameter              & Value   & Notes \\
\hline
\multicolumn{2}{l}{{\bf Stellar Parameters}} \\
\teff (K)                 & $\sys{steff} \pm \sys{steff_err}$ & A \\
\logg (dex)               & $\sys{slogg} \pm \sys{slogg_err}$ & A \\
\fe (dex)                 & $\sys{smet} \pm \sys{smet_err}$   & A \\
$K$ (mag)                 & $11.2 \pm 0.03$                   & B \\
$\pi_{\star}$ (mas)       & $3.42 \pm 0.06$                   & C \\
\multicolumn{2}{l}{{\bf Photodynamical Analysis}} \\
\Mstar (\Msun)            & \sys{pd-mstar_fmt}       & D,E \\
\Rstar (\Rsun)            & \sys{pd-rstar_fmt}       & D,E \\
$q_1$                     & \sys{pd-c1_fmt}          & D \\
$q_2$                     & \sys{pd-c2_fmt}          & D \\
$P_b$ (days)              & \sys{pd-per2_fmt}        & D \\
$\tc{b}$ (BJD$-$2454833)  & \sys{pd-tc2_fmt}         & D \\
$\secosw{b}$              & \sys{pd-secosw2_fmt}     & D \\
$\sesinw{b}$              & \sys{pd-sesinw2_fmt}     & D \\
$\inc{b}$ (deg)           & \sys{pd-inc2_fmt}        & D \\
$\Omega_b$ (deg)          & 0 (fixed)                & D \\
$\rrat{b}$                & \sys{pd-ror2_fmt}        & D \\
$M_{p,b}$ (\Me)           & \sys{pd-masseprec2_fmt}  & D,F \\
$P_c$ (days)              & \sys{pd-per3_fmt}        & D \\
$\tc{c}$ (BJD$-$2454833)  & \sys{pd-tc3_fmt}         & D \\
$\secosw{c}$              & \sys{pd-secosw3_fmt}     & D \\
$\sesinw{c}$              & \sys{pd-sesinw3_fmt}     & D \\
$\inc{c}$ (deg)           & \sys{pd-inc3_fmt}        & D \\
$\Omega_c$ (deg)          & \sys{pd-Omega3_fmt}      & D \\
$\rrat{c}$                & \sys{pd-ror3_fmt}        & D \\
$M_{p,c}$ (\Me)           & \sys{pd-masse3_fmt}      & D,F \\
$P_d$ (days)              & \sys{pd-per1_fmt}        & D \\
$\tc{d}$ (BJD$-$2454833)  & \sys{pd-tc1_fmt}         & D \\
$\secosw{d}$              & 0 (fixed)                & D \\
$\sesinw{d}$              & 0 (fixed)                & D \\
$\inc{d}$ (deg)           & \sys{pd-inc1_fmt}        & D \\
$\Omega_d$ (deg)          & 0 (fixed)                & D \\
$\rrat{d}$                & \sys{pd-ror1_fmt}        & D \\
$M_{p,d}$ (\Me)           & <10                      & D,F \\
\multicolumn{2}{l}{{\bf Derived Parameters}} \\
$R_{p,b}$ (\Re)           & \sys{pd-prad2_fmt}       & G \\
$R_{p,d}$ (\Re)           & \sys{pd-prad1_fmt}       & G \\
$R_{p,c}$ (\Re)           & \sys{pd-prad3_fmt}       & G \\
$\ecc{b}$                 & \sys{pd-e2_fmt}          & G \\
$\ecc{c}$                 & \sys{pd-e3_fmt}          & G \\
$\Delta \omega$ (deg)     & \sys{pd-omegadiffdeg_fmt}& G \\
$f_\mathrm{env,b}$ (\%)   & \sys{lopez-fenv2_fmt}    & H \\
$f_\mathrm{env,c}$ (\%)   & \sys{lopez-fenv3_fmt}    & H \\
$M_\mathrm{core,b}$ (\Me)  & \sys{lopez-mcore2_fmt}  & H \\
$M_\mathrm{core,c}$ (\Me)  & \sys{lopez-mcore3_fmt}  & H \\
\hline
\\[-6ex]
\end{tabular}
\end{center}
\tablecomments{A: \cite{Brewer16}. B: 2MASS \citep{Skrutskie06}. C: {\em Gaia} DR2 \citep{Gaia18}. D: Input parameters into photodynamical model, see Section~\ref{sec:photodyn}. E: We imposed Gaussian priors on $\Mstar$ and $\Rstar$ using the methodology described in \cite{Fulton18b} that incorporated A, B, and C. F: Based on our RV analysis (Section~\ref{sec:rv-keplerian}), we imposed the following Gaussian priors on planet masses: $M_{p,b}$ = \sys{rv-mpsini1_fmt}~\Me, $M_{p,c}$ = $0\pm10~\Me$, $M_{p,d}$ = $0.0\pm3.2$~\Me. G: Derived from the posterior samples of D. H: Derived from the planet mass and radius constraints along with the core-envelope models of \cite{Lopez14}. See Section~\ref{sec:core-envelope} for further details.}
\label{tab:sys}
\end{table}

\begin{figure*}
\centering
\includegraphics[width=1.0\textwidth]{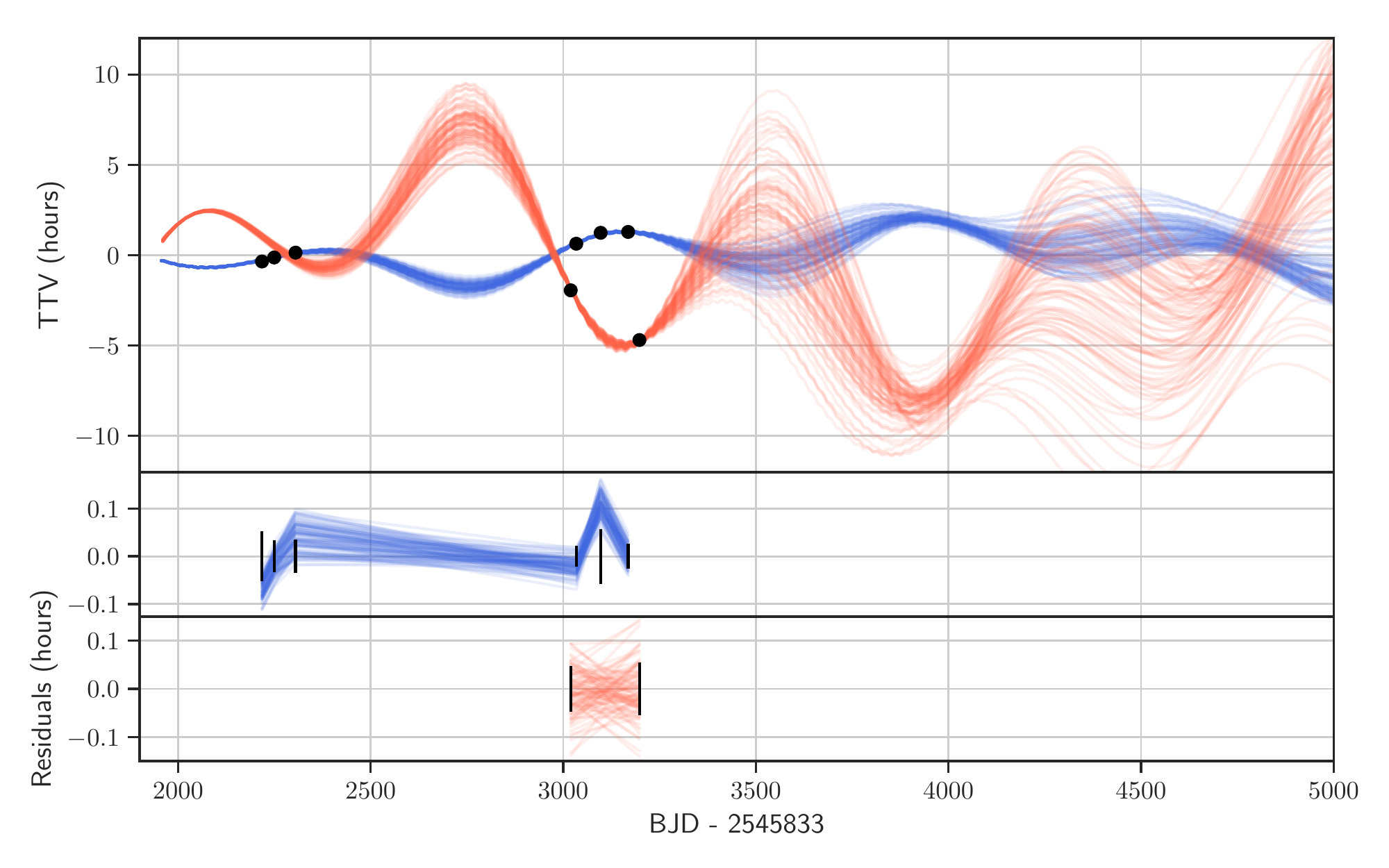}
\caption{Top: black points show measured transit times with respect to a reference linear ephemeris. Blue and orange lines show transit times of K2-19b and c, respectively, computed from 100 draws from our MCMC chains (see Section~\ref{sec:photodyn}). We do not show times from the \ktwo epoch ($t$ = 1980--2060~days) because we model the flux timeseries directly. The lines in the bottom panels represent the residuals to the predicted transit times and the formal timing uncertainties. Most of the model draws are within 2$\sigma$ of the measured transit times and indicate good agreement between data and model.\label{fig:photodyn-samples-times}}
\end{figure*}

\begin{figure}
\centering
\includegraphics[width=0.49\textwidth]{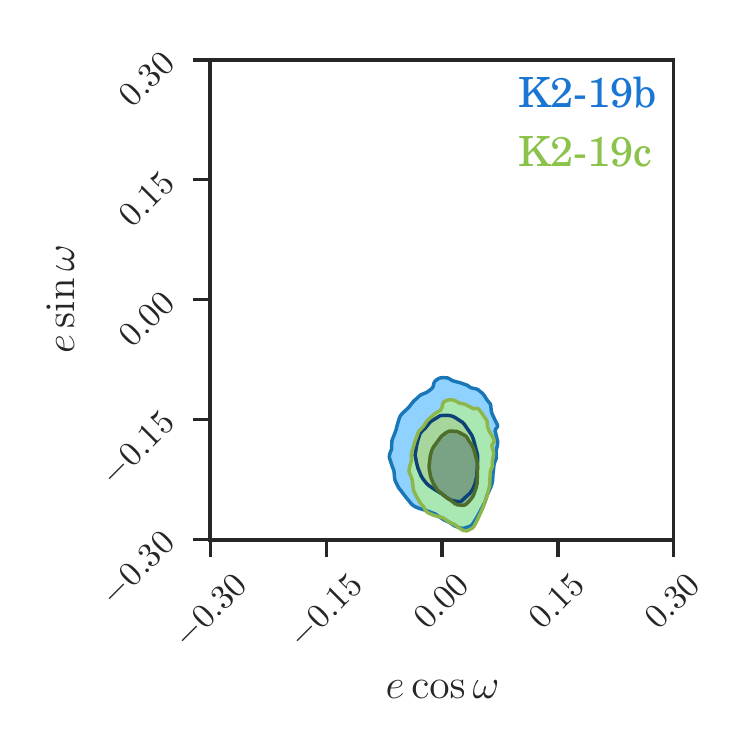}
\caption{2D joint posterior of $e \cos \omega$ and $e \sin \omega$ for K2-19b (blue) and K2-19c (green); the contours show 1 and 2 sigma levels. The eccentricities of both planets (i.e. the radial distance from the origin) are non-zero and are consistent to within errors, $\ecc{b}$ = \sys{pd-e2_fmt} and $\ecc{c}$ = \sys{pd-e3_fmt}. The planets have well-aligned apsides with $\Delta \omega$  = \sys{pd-omegadiffdeg_fmt} (deg). \label{fig:corner-ecc}}
\end{figure}

\section{Core/Envelope Structure}
\label{sec:core-envelope}
Here, we examine the K2-19 planets in the context of other known exoplanets. Figure~\ref{fig:context} shows a mass-radius diagram constructed from the NASA Exoplanet Archive \citep{Akeson13}. Our $\sim$5\% mass measurements are among the most precise for any sub-Jovian size planet. Mass and radius reflect a planet's core/envelope distribution. K2-19 are both ``sub-Saturns,'' which we define as planets with \Rp = 4--8~\Re. The bulk composition of sub-Saturns may be well-approximated by a two-component model consisting of a high density core and a H/He envelope of solar composition \citep{Lopez14,Petigura16}. For sub-Saturns, their total size is determined largely by their envelope fraction $\fenv = M_\mathrm{env} / \Mp$, and thus changes in the detailed core composition weakly affect the total size.

\cite{Lopez14} computed planet radii over a grid of $\Mp$, $\fenv$, age, and incident flux \Sinc. As a point of reference, we show the mass-radius relationship for these models at several values of \fenv in Figure~\ref{fig:context-massradius}.%
\footnote{Formally, we set age = 5 Gyr and \Sinc = 80~\Se in order to plot single lines, but we note that these curves are nearly overlapping at low \Sinc and late times.}
Both K2-19b and c require volumetrically significant envelopes to explain their masses and sizes. Following \cite{Petigura17a}, we derived core masses and envelope fractions for these planets by interpolating over the \cite{Lopez14} model grid. K2-19c has a core mass of \sys{lopez-mcore3_fmt}~\Me and is \sys{lopez-fenv3_fmt}\% envelope by mass, while K2-19b has a core mass of \sys{lopez-mcore2_fmt}~\Me and is \sys{lopez-fenv2_fmt}\% envelope by mass.

\cite{Petigura17a} compiled a sample of 23 sub-Saturns with well-measured masses and radii to examine trends within this population. One trend is that sub-Saturns have a range of envelope fractions, and that range broadens with decreasing equilibrium temperature. This broadening is likely due to the decreasing importance of photoevaporation at lower \teq. The K2-19 planets have intermediate \teq of $\sim$800~K and span the full range of \fenv. 

\cite{Petigura17a} also noted a positive correlation between the host star metallicity and the total mass of sub-Saturns. As intermediate mass sub-Saturns around a near solar-metallicity star, the K2-19 planets also conform to this trend. The emerging \Mp--\fe correlation may point to metallicity dependent effects in the growth of cores and/or accretion of gas from the protoplanetary disk. However, an expanded sample size is needed to more thoroughly assess the significance of this correlation and possible dependencies on quantities like stellar mass, which is covariant with metallicity.

With \fenv = \sys{lopez-fenv2_fmt}\%, K2-19b is one of the most envelope-rich sub-Saturns known. Its envelope fraction is nearly as high as K2-24c with \fenv = $52^{+5}_{-3}\%$ \citep{Petigura18c}. Like K2-24c, K2-19b presents an intriguing challenge to traditional core-accretion theory. As a point of reference, in the canonical core accretion models of \cite{Pollack96}, Saturn forms first as a $\approx$12~\Me core that accretes H/He from the protoplanetary disk. At the crossover mass (i.e. when $\menv \approx \mcore$ or when $\fenv \approx 50\%$), runaway accretion begins and Saturn quickly grows to its final mass.

One could attempt to resolve the $\fenv \approx 50\%$ problem by imagining that the disk dissipated right as K2-19b approached the runaway phase. While this scenario is impossible to rule out, it requires special timing of planet formation and is thus a priori unlikely. More likely, the inferred structure of K2-19b points to an incomplete understanding of core-nucleated accretion and motivates further theoretical explanations of planet conglomeration in the sub-Saturn mass regime.

\section{Mean-Motion Resonance}
\label{sec:mmr}
K2-19b and c are clearly {\em near} the 3:2 mean-motion resonance, but are they actually {\em in} resonance? Resonance requires the libration of a resonant angle, e.g.,
\[
\phi = 3 \lambda_c - 2 \lambda_b - \varpi,
\]
where $\lambda$ is the mean longitude and $\varpi$ is the longitude of periastron for either planet b or c. Librating angles are confined to a particular range while circulating angles sweep out all values between 0 and $2\pi$. If $\phi$ is librating,
\[
\langle\dot{\phi}\rangle  = 3 n_c - 2 n_b - \dot{\varpi} = 0. 
\]

We simulated the plausible long-term evolution of K2-19b and c by taking 100 draws from the posterior samples from Section~\ref{sec:photodyn} and evolving them for 50~years using the IAS15 $N$-body integrator included in the REBOUND package \citep{Rein12,Rein15}.

Our integrations all revealed the same qualitative apsidal outcome: circulation rather than libration of $\langle\phi\rangle$. In Figure~\ref{fig:rebound}, we show the evolution of $\dot{\phi}$ for a representative simulation. The quantity $3 n_c - 2 n_b$ has a time average of $-0.003$~rad/day, much larger than $\dot{\varpi_b}$ or $\dot{\varpi_c}$. Instead, the planet eccentricities evolve secularly over a period of roughly six years while the apsides remain aligned.

In our simulations we did not include precession from general relativity or the quadrupole field due to K2-19d. Here, we justify these approximations. Planet b experiences apsidal precession due to an effective quadrupole moment from planet d. The rate of this precession is given by
\[
\dot{\omega}_{J2} = 3 n_b J_2 \left(\frac{a_d}{a_b}\right)^2
\]
where 
\[
J_2 = \frac{1}{2}\frac{m_d}{M_\star}.
\]
We find that 
\[
\tau_{J2} = 2 \pi /  \dot{\omega}_{J2} \approx 2\times10^4\, \mathrm{yr}.
\]
K2-19b also experiences apsidal precision due to GR with a rate of 
\[
\dot{\omega}_{GR} = 3 n_b \frac{G M_\star}{a_b c^2}
\]
so that
\[
\tau_{GR} = 2 \pi /  \dot{\omega}_{GR} \approx 6\times10^4\, \mathrm{yr}.
\]
Because $\tau_{GR}$ and $\tau_{J2}$ are much longer than the secular eccentricity oscillations, we are justified in neglecting their effects above.

\section{Formation}
\label{sec:formation}

An intriguing aspect of the K2-19 system is that both the physical and orbital characteristics of planets b and c are peculiar, especially when viewed against the backdrop of other well-characterized planetary systems, including our own. In particular, from the perspective of conventional planet formation theory \citep{Armitage10}, the inferred properties of the K2-19 planets present a formidable challenge. As already mentioned above, the near-unity envelope-to-core mass fraction of K2-19c is not a natural outcome of core-nucelated accretion model of planet formation \citep{Pollack96,Hubickyj05}. However, even if we ignore the physical structure of these planets altogether, their orbital architecture lies in sharp contrast with with theoretical expectations \citep{Kley12}.

The most noteworthy feature of the K2-19bc pair is their proximity to exact 3:2 mean motion commensurability.  In general, orbital resonances have long been recognized as an aftereffect of convergent orbital migration \citep{Tanaka02,Bitsch15}. Furthermore, theoretical treatment of migration predicts that planets as massive as K2-19b and c should have readily experienced disk-driven orbital decay. Therefore, it is not unreasonable to anticipate a distinctly resonant present-day architecture of K2-19 that could in turn be attributed to a migratory origin. Moreover, coupled with long-range migration, resonant interactions are well-known to adiabatically excite the orbital eccentricities of the constituent planets (see, e.g., \citealt{Peale86,Malhotra95,Lee02}), and our photodynamical model revealed significant eccentricities of $e \approx 0.2$. Nevertheless, as we showed in Section~\ref{sec:mmr}, the system is incompatible with mean-motion resonance, and thus the entire aforementioned narrative.

Both the values of the eccentricities themselves, as well as the apsidal orientations of the orbits are contradictory to those that would have been sculpted by convergent migration. More specifically, within the framework of the standard resonance capture scenario, orbital eccentricities are determined by a balance between adiabatic excitation that arises from convergent orbital evolution and disk-driven eccentricity damping. Quantitatively, this balance yields eccentricities of $e \sim h / r \sim 0.05$, where $h$ is the disk scale height and $r$ is the distance to the host star \citep{Pichierri18}.

However, the inferred eccentricities of K2-19b and c exceed this characteristic value by a factor of a few. More dramatically, a clear consequence of adiabatic resonance capture is the anti-alignment of planetary apsidal lines, such that $\Delta \varpi \approx 180$~deg \citep{Batygin13a}. Instead, in this system, the data clearly points to apsidal alignment, characterized by $\Delta \varpi \approx 0$~deg. It is this requirement for the periapse alignment that prevents us from finding a suitable resonant solution for the planetary orbits.

To elaborate on apsidal alignment further, we note that stable resonant equilibria that exist far away from $\Delta \varpi \approx 180$~deg are indeed possible at sufficiently high eccentricities \citep{Beauge06}. In an effort to consider this possibility for K2-19, we carried out an N-body numerical experiment, simulating the convergent migration and subsequent resonant locking of K2-19b and c. In particular, we initialized both planets on circular orbits, at a period ratio 20\% outside of nominal 3:2 commensurability and computed the orbital evolution resulting from mutual gravitational perturbations as well as a fictitious force designed to mimic planet-disk interactions. The integration was carried out using the Bulirsch-Stoer algorithm \citep{Press92}, with an accuracy parameter set to $10^{-10}$.

We adopted the model disk acceleration formulae spelled out in \cite{Papaloizou00}, setting the convergent migration timescale $\tau_a=2\times10^4$~yr. While our choice of $\tau_a$ was arbitrary, the resulting evolution is adiabatic and thus insensitive to the adopted $\tau_a$ \citep{Henrard82}. To prevent the system from equilibrating in resonance with low eccentricities (e.g. \citealt{Pichierri17}), we unphysically set the timescale for eccentricity damping to $\tau_e=\infty$, such that disk-driven convergent migration resulted in continued adiabatic enhancement of the planetary eccentricities once a resonant coupling was established \citep{Lee04}.

The initial results of our simulations followed a familiar pattern: the planets migrated convergently, were captured into the 3:2 mean-motion resonance, and developed finite eccentricities while locked into strict apsidal anti-alignment with $\Delta\varpi=180\deg$. Once the planetary eccentricities reached sufficiently large values, however, we observed deviations away from exact apsidal anti-alignment. Nevertheless, we found that  in order to attain $\Delta\varpi$ even remotely close to zero, unreasonably large eccentricities were required. For example, a resonant equilibrium at $\Delta\varpi \sim 60 \deg$ requires $e > 0.8$ for both planets. Thus, our results show that although asymmetric equilibria can follow after capture into mean-motion resonance, the required eccentricities are simply too high to be observationally permissible. Indeed, resonant coupling appears to be strictly ruled out by the available data.%
\footnote{As a corollary, we note that orbits which originate in resonance can be driven out of exact commensurability while maintaining libration of resonant angles by long-term energy dissipation \citep{Lithwick12b,Batygin13b}. This scenario, however is only relevant to systems with vanishingly low eccentricities and period ratios well outside of the nominal resonance width, both of which are not satisfied in K2-19.}

For completeness, we can also speculate regarding an alternative mechanism for finite eccentricity excitation: mean-motion resonance crossing due to divergent migration. In this scenario, planets start out interior to a resonant period ratio and cross a commensurability, which results in a non-capturing encounter with the resonant separatrix. This yields an impulsive excitation of the planetary eccentricities. For example, models of the early solar system by  \cite{Tsiganis05}, planetesimal scattering by Jupiter and Saturn leads to divergent migration, and the crossing of the 2:1 resonance excites eccentricities of $\approx$5--10\%. This scenario, however, also yields strict apsidal anti-alignment after the encounter \citep{Batygin13a} and is therefore also ruled out by the observations. 

We conclude this section with a brief remark on dynamical stability and its relationship to the observed orbital architecture of the K2-19 system. A trivial examination of the derived orbital elements illustrates that this systems is strongly AMD-unstable \citep{Petit18}. So how is the stability of these planets ensured? It is well known that highly eccentric planets or satellites locked into orbital commensurabilities often derive long-term orbital stability from the resonant phase-protection mechanism. As we have demonstrated above, however, in the case of K2-19b and c, libration of resonant angles appears to be forbidden by the observational data. Instead, the planets around K2-19 appear to be protected from close encounters primarily by the fact that the orbits have persistently co-linear apses and are therefore geometrically nested. While this configuration is indeed long-term stable, the dynamical genesis of this orbital configuration remains elusive.

\begin{figure}
\centering
\includegraphics[width=0.5\textwidth]{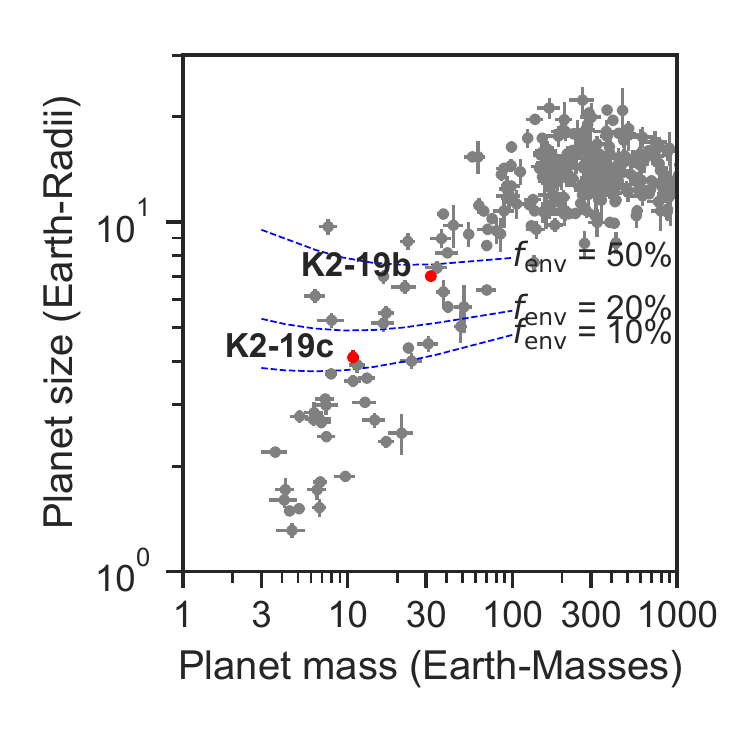}
\caption{The K2-19 planets viewed in context with other exoplanets. Gray points show the masses and radii of exoplanets where mass is measured to 25\% or better. The K2-19 planets are shown in red and the uncertainties are comparable to the point size. The blue lines show mass-radius relationships for model planets having an Earth-composition core and various envelope fractions of H/He, \fenv = $M_\mathrm{env} / \Mp$.\label{fig:context-massradius}}
\end{figure}

\begin{figure*}
\centering
\hspace{-0.5cm}
\includegraphics[width=0.45\textwidth]{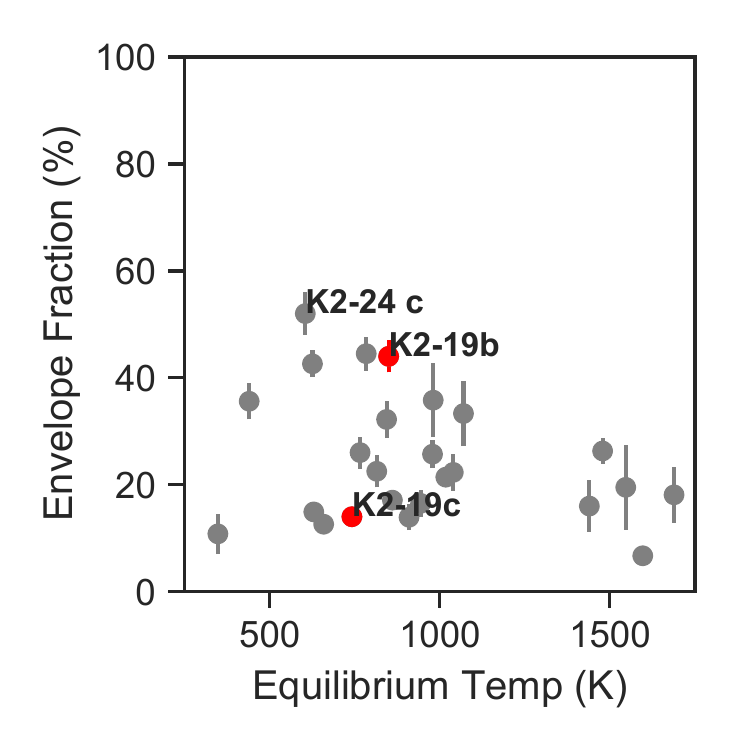}
\hspace{-0.5cm}
\includegraphics[width=0.45\textwidth]{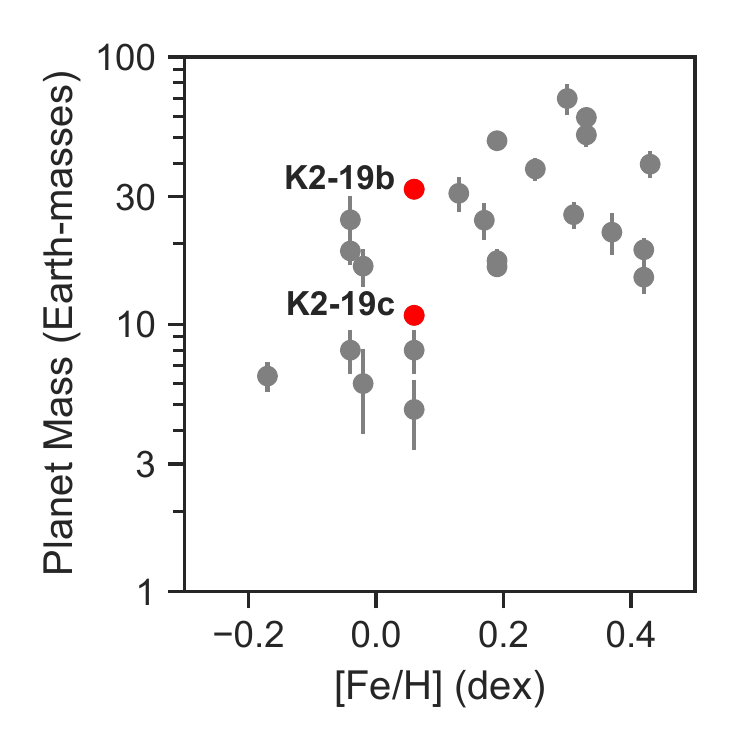}
\caption{The K2-19 planets viewed in context amongst other sub-Saturns (\Rp = 4--8~\Re). Left: Envelope fraction vs. equilibrium temperature. Gray points are drawn from \cite{Petigura17a}. The K2-19 planets straddle the range of observed envelope fractions. K2-19b resides near the upper envelope of the \fenv distribution, which broadens toward lower \teq. Right: Same sample as left, but showing planet mass vs. host star metallicity. There is a positive correlation between \fe and \Mp, with significant scatter. The K2-19 planets conform to this trend.\label{fig:context}}
\end{figure*}

\begin{figure*}
\centering
\includegraphics[width=1\textwidth]{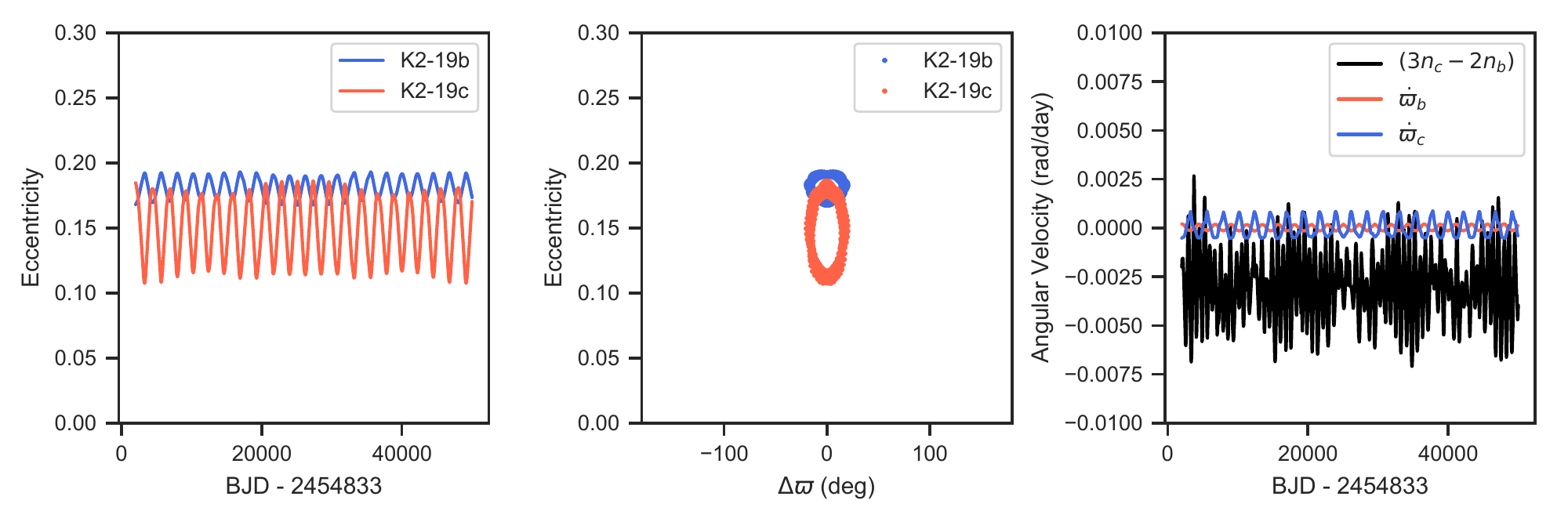}
\caption{The dynamical evolution of a representative solution from our photodynamical model, based on a 50 year $N$-body integration. Left: Eccentricity as a function of time. The planets exchange eccentricity over a secular timescale of $\sim$6 years. Middle: same simulation as left, but with $\Delta \varpi$ on the x-axis. The planets precess together and retain apsidal alignment. Right: Several angular velocities relevant to the planet's resonant state. Because the quantity $3 n_c - 2 n_b - \dot{\varpi}$ does not have a time average of zero for either $\dot{\varpi}_b$ or $\dot{\varpi}_c$, the planets are not in the 3:2 mean motion resonance. Instead, the resonant angles circulate at a rate of $\sim$1 radian per year.\label{fig:rebound}}
\end{figure*}

\vspace{1cm}

\section{Conclusions}
\label{sec:conclusions}

The K2-19 system offers a sharp contrast to the architecture and physical properties of the solar system planets. In the solar system, not a single planet resides interior to Mercury ($P = 88$~d), while for K2-19 there are (at least) three planets with $P < 12$~d. K2-19c straddles a gap in the size distribution of solar system planets between the ice giants and Jovians. Finally, no pair of major solar system planets resides so close to mean-motion resonance, although numerous Kuiper belt objects are in resonance with Neptune, of which Pluto is the prototypical example.

The planets orbiting K2-19 are also unusual compared to typical extrasolar planets. Highly irradiated planets between size of Neptune and Saturn are rare: \cite{Petigura18b} performed a demographic analysis of GK stars observed by \Kepler and found 0.36 planets per 100 stars with \Rp = 4--8~\Re and $P < 10$~d. In addition, such proximity to resonance is not a common feature of extrasolar planets; to first order, planet period ratios are uniformly distributed \citep{Lissauer11}. 

Motivated by the unusual characteristics of the K2-19 planets, our team collected RVs with Keck/HIRES and additional photometry from \spitzer and LCO. The RV dataset was sufficient to detect the reflex motion due to K2-19b at $7\sigma$. However, the RVs alone were insufficient to detect K2-19c due to its lower mass. Quasiperiodic RV variability due to spots of $\approx$7~\ms limited the sensitivity the RV dataset. Spot contrasts are smaller at redder wavelengths, and K2-19 would benefit from RV monitoring in the NIR by instruments such as IRD \citep{Kotani18}.

The high precision of the \ktwo and \spitzer photometry combined with our multi-year time baseline provided much more stringent constraints on the physical and orbital properties of the planets. We measured the masses of both K2-19b and c to $\approx$5\%, which are among the most precise of any sub-Jovian exoplanet. Our mass and radius measurements provided a window into the core-envelope structure of these planets. We found that K2-19c is roughly 15\% envelope by mass, while K2-19b is nearly 50\%---close to the canonical cross-over mass leading to runaway accretion \citep{Pollack96}. These planets contribute to an emerging picture of planets between size of Neptune and Saturn: where cores of a given mass exhibit a wide diversity of envelope fractions and where that diversity grows with decreasing irradiation (see Figure~\ref{fig:context}).

Through our photodynamical analysis, we found that these planets have moderate eccentricities of $\approx$0.2 and aligned apsides. The planets are experiencing rapid secular eccentricity oscillations with a $\approx$6~yr timecale, but the system is currently not in mean-motion resonance. Moreover, the system's present configuration presents a challenge to formation pathways that involve mean-motion resonance in the past. Scenarios where the system passes through the 3:2 resonance from above or below predict anti-aligned apsides, which are ruled out by the data. Future photometric or RV monitoring would shed additional light on this enigmatic system.
 
\acknowledgements
We thank the anonymous reviewer for helpful suggestions that improved this manuscript.

E.A.P. acknowledges support for this work by NASA through the NASA Hubble Fellowship grant HST-HF2-51417 awarded by the Space Telescope Science Institute, which is operated by the Association of Universities for Research in Astronomy, Inc., for NASA, under contract NAS5-26555. S.C.C.B. acknowledges support from Funda\c{c}\~{a}o para a Ci\^{e}ncia e a Tecnologia (FCT) through Investigador FCT contract IF/01312/2014/CP1215/CT0004 and by FEDER through COMPETE2020 and POCI in the framework of the project POCI-01-0145-FEDER-028953.

This work made use of NASA's Astrophysics Data System Bibliographic Services. 

This work is based in part on observations made with the Spitzer Space Telescope, which is operated by the Jet Propulsion Laboratory, California Institute of Technology under a contract with NASA. Support for this work was provided by NASA through an award issued by JPL/Caltech. This work makes use of observations from the LCOGT network.

Some of the data presented herein were obtained at the W.\ M.\ Keck Observatory, which is operated as a scientific partnership among the California Institute of Technology, the University of California, and NASA. The authors wish to recognize and acknowledge the very significant cultural role and reverence that the summit of Maunakea has always had within the indigenous Hawaiian community.  We are most fortunate to have the opportunity to conduct observations from this mountain.

\facilities{Kepler, Spitzer, Keck:I (HIRES), LCOGT}

\software 
{batman} \citep{Kreidberg15}, emcee \citep{Foreman-Mackey13}, Phodymm \citep{Mills16}, REBOUND \citep{Rein12}, EVEREST2.0 \citep{Luger17}, RadVel \citep{Fulton18a}.

\bibliography{manuscript.bib}

\appendix
\section{Photodynamical Model}
\label{sec:appendix}

Here, we include some supplemental information regarding our photodynamical model described in Section~\ref{sec:photodyn}. Table~\ref{tab:transit-times-predict} lists the predicted transit times and uncertainties for K2-19b and c up to 2029. Figure~\ref{fig:corner} shows the 2D joint posteriors of all parameters included in our photodynamical model.

We highlight the covariances between mass and eccentricity in Figure~\ref{fig:corner-mass-ecc}. The masses of planet b and c are correlated because the amplitudes of near-resonant TTVs constrain planet mass ratios \citep{Lithwick12a}. However, the RVs and higher order TTV terms (i.e. chopping) constrain the individual masses directly \citep{Deck15b}. Figure~\ref{fig:corner-mass-ecc} also illustrates a positive correlation between $e_b \cos \omega_b$ and $e_c \cos \omega_c$ and between  $e_b \sin \omega_b$ and $e_c \sin \omega_c$. This is another common feature of near-resonant systems: the TTV amplitude and phase encodes linear combinations of $e \cos \omega$ and $e \sin \omega$ \citep{Lithwick12a}.

\begin{deluxetable}{llrrr}[h]
\tablecaption{Predicted Transit Times\label{tab:transit-times-predict}}
\tablehead{
  \colhead{Planet} & 
  \colhead{$i$} & 
  \colhead{UTC date} & 
  \colhead{$T_{c}$} & 
  \colhead{$\sigma(T_c)$} \\ 
  \colhead{} & 
  \colhead{} & 
  \colhead{} & 
  \colhead{days} & 
  \colhead{days} 
}
\startdata
\input{tab_transit-times-predict-stub1.tex}
\hline
\input{tab_transit-times-predict-stub2.tex}
\enddata
\tablecomments{Predicted transit times for K2-19b and c, where $i$, is an index that labels individual transits. Times are given in $\mathrm{BJD}_\mathrm{TBD} - 2454833$. Table~\ref{tab:transit-times} is published in its entirety in the machine-readable format. A portion is shown here for guidance regarding its form and content.}
\end{deluxetable}

\begin{figure*}[h!]
\centering
\includegraphics[width=1\textwidth,trim={1cm 6cm 1cm 0cm}]{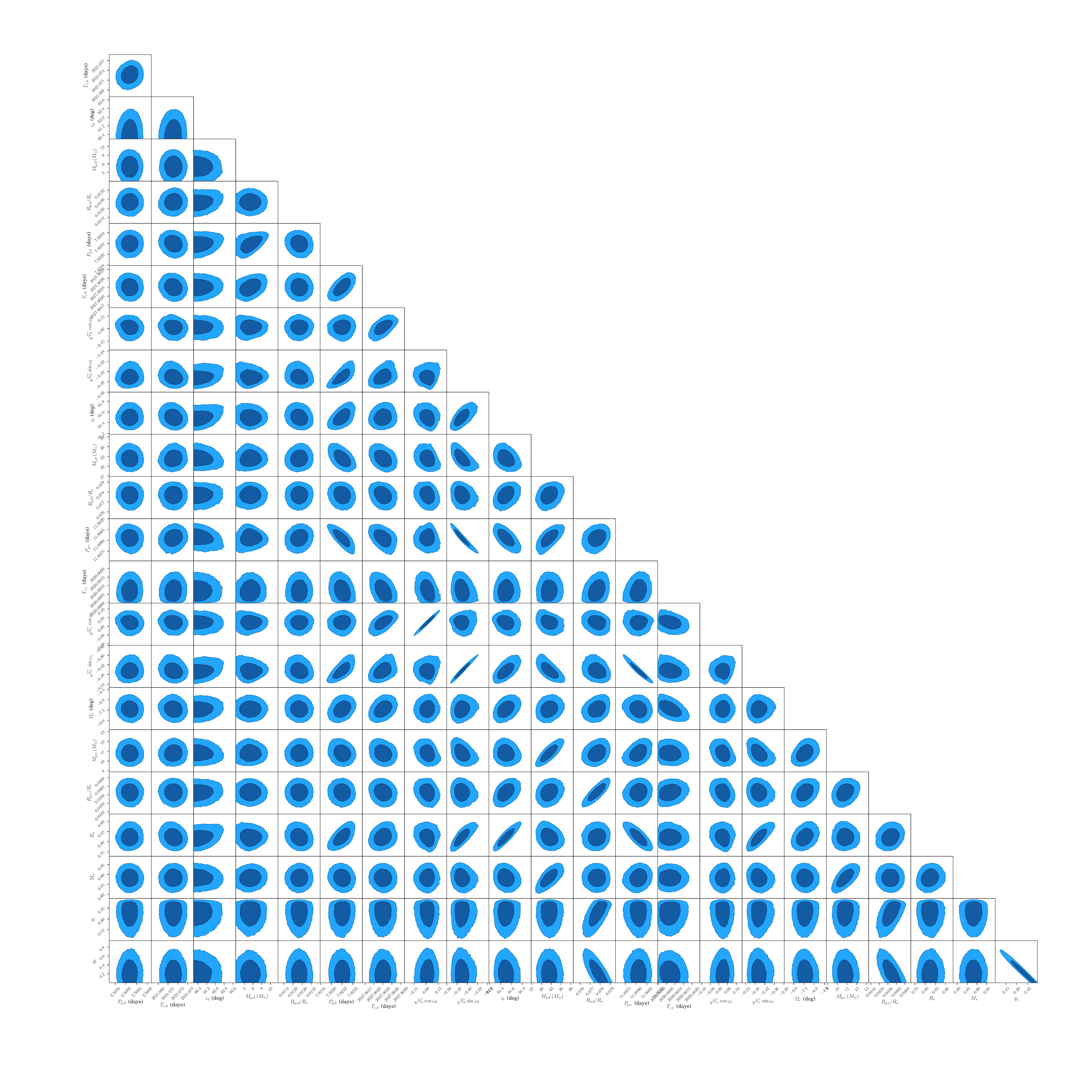}
\caption{2D joint posterior probability distributions for our photodynamical model (Section~\ref{sec:photodyn}). The dark and light regions show 1 and 2 sigma contours, respectively.\label{fig:corner}}
\end{figure*}

\begin{figure*}
\centering
\includegraphics[trim={0.3cm 0.3cm 0.3cm 0.3cm},width=0.32\textwidth]{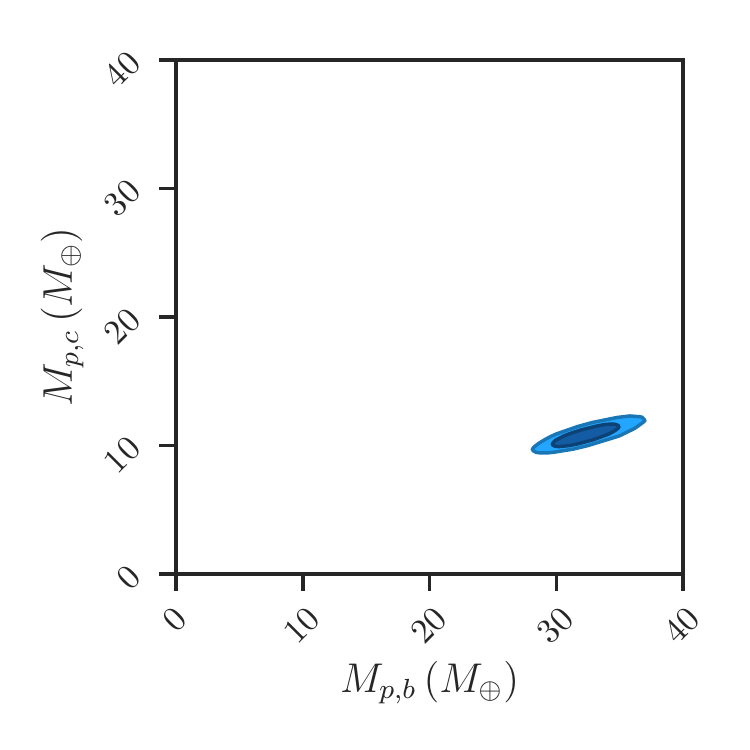}
\includegraphics[trim={0.3cm 0.3cm 0.3cm 0.3cm},width=0.32\textwidth]{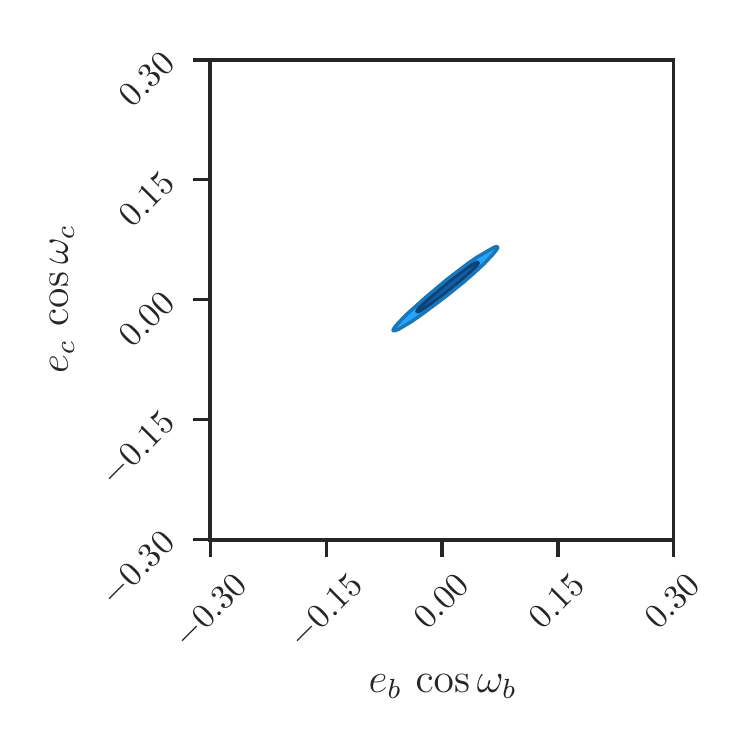}
\includegraphics[trim={0.3cm 0.3cm 0.3cm 0.3cm},width=0.32\textwidth]{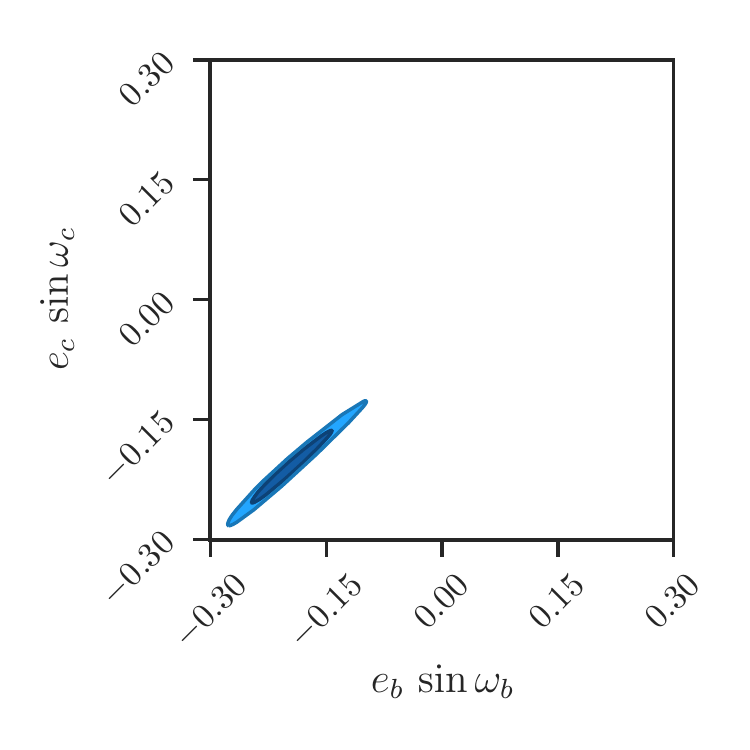}
\caption{Same as Figure~\ref{fig:corner}, but highlighting several noteworthy covariances between planet masses and planet eccentricities. Left panel: Constraint on $M_{p,b}$ and $M_{p,c}$. The covariance between the planet masses is typical of TTV analyses which tend to provide smaller fractional uncertainties on mass ratios than on individual masses. Middle panel: same as left but for $e_b \cos \omega_b$ and $e_c \cos \omega_c$. The covariance results from the fact that TTVs constrain linear combinations of the eccentricity vectors. Right panel: same as middle but for  $e_b \sin \omega_b$ and $e_c \sin \omega_c$.\label{fig:corner-mass-ecc}}
\end{figure*}

\end{document}

%% file: val_stat.tex
\newcommand{\stat}[1]{%%
\IfEqCase{#1}{%
{rv-n}{51}%
{rv-start}{2015-02-05}%
{rv-stop}{2017-12-26}%
{rv-errvel-min}{1.9}%
{rv-errvel-max}{3.8}%
{rv-errvel-med}{2.2}%
{rv-snr-min}{53}%
{rv-snr-max}{108}%
{rv-snr-med}{91}%
}[XX]%
}

%% file: val_sys.tex
\newcommand{\sys}[1]{%%
\IfEqCase{#1}{%
{steff}{5322}%
{steff_err}{100}%
{slogg}{4.51}%
{slogg_err}{0.08}%
{smet}{0.06}%
{smet_err}{0.05}%
{smass}{0.88}%
{smass_err}{0.03}%
{srad}{0.82}%
{srad_err}{0.04}%
{phot-gp-eta1}{0.4}%
{phot-gp-eta1_err}{0.1}%
{phot-gp-eta2}{34}%
{phot-gp-eta2_err}{5}%
{phot-gp-eta3}{20.4}%
{phot-gp-eta3_err}{0.3}%
{phot-gp-eta4}{0.49}%
{phot-gp-eta4_err}{0.06}%
{rv-per1}{7.92075}%
{rv-per2}{11.89804}%
{rv-per3}{2.50856}%
{rv-tc1}{1980.379}%
{rv-tc2}{1984.305}%
{rv-tc3}{1975.921}%
{rv-k1}{11.4}%
{rv-k1_err1}{1.7}%
{rv-k1_err2}{-1.7}%
{rv-k1_fmt}{$11.4 \pm 1.7$}%
{rv-k2}{0.4}%
{rv-k2_err1}{1.5}%
{rv-k2_err2}{-1.5}%
{rv-k2_fmt}{$0.4 \pm 1.5$}%
{rv-k3}{-0.2}%
{rv-k3_err1}{1.0}%
{rv-k3_err2}{-1.0}%
{rv-k3_fmt}{$-0.2 \pm 1.0$}%
{rv-mpsini1}{33}%
{rv-mpsini1_err1}{5}%
{rv-mpsini1_err2}{-5}%
{rv-mpsini1_fmt}{$33 \pm 5$}%
{rv-mpsini2}{1}%
{rv-mpsini2_err1}{5}%
{rv-mpsini2_err2}{-5}%
{rv-mpsini2_fmt}{$1 \pm 5$}%
{rv-mpsini3}{-0}%
{rv-mpsini3_err1}{2}%
{rv-mpsini3_err2}{-2}%
{rv-mpsini3_fmt}{$-0 \pm 2$}%
{rv-musini1}{111}%
{rv-musini1_err1}{17}%
{rv-musini1_err2}{-17}%
{rv-musini1_fmt}{$111 \pm 17$}%
{rv-musini2}{5}%
{rv-musini2_err1}{17}%
{rv-musini2_err2}{-17}%
{rv-musini2_fmt}{$5 \pm 17$}%
{rv-musini3}{-1.1}%
{rv-musini3_err1}{6.7}%
{rv-musini3_err2}{-6.7}%
{rv-musini3_fmt}{$-1.1 \pm 6.7$}%
{rv-gamma_j}{-2}%
{rv-gamma_j_err1}{2}%
{rv-gamma_j_err2}{-2}%
{rv-gamma_j_fmt}{$-2 \pm 2$}%
{rv-dvdt}{5.9}%
{rv-dvdt_err1}{2.4}%
{rv-dvdt_err2}{-2.4}%
{rv-dvdt_fmt}{$5.9 \pm 2.4$}%
{rv-gp_amp}{7.4}%
{rv-gp_amp_err1}{2.2}%
{rv-gp_amp_err2}{-2.2}%
{rv-gp_amp_fmt}{$7.4 \pm 2.2$}%
{rv-gp_explength}{35.0}%
{rv-gp_explength_err1}{4.7}%
{rv-gp_explength_err2}{-4.7}%
{rv-gp_explength_fmt}{$35.0 \pm 4.7$}%
{rv-gp_per}{20.3}%
{rv-gp_per_err1}{0.3}%
{rv-gp_per_err2}{-0.3}%
{rv-gp_per_fmt}{$20.3 \pm 0.3$}%
{rv-gp_perlength}{0.46}%
{rv-gp_perlength_err1}{0.07}%
{rv-gp_perlength_err2}{-0.07}%
{rv-gp_perlength_fmt}{$0.46 \pm 0.07$}%
{rv-mpsini2_p95}{10.2}%
{rv-mpsini3_p95}{3.5}%
{pd-mstar}{0.88}%
{pd-mstar_err1}{0.03}%
{pd-mstar_err2}{-0.03}%
{pd-mstar_fmt}{$0.88 \pm 0.03$}%
{pd-rstar}{0.82}%
{pd-rstar_err1}{0.03}%
{pd-rstar_err2}{-0.03}%
{pd-rstar_fmt}{$0.82 \pm 0.03$}%
{pd-masse1}{5.1}%
{pd-masse1_err1}{2.4}%
{pd-masse1_err2}{-2.4}%
{pd-masse1_fmt}{$5.1 \pm 2.4$}%
{pd-masse2}{32}%
{pd-masse2_err1}{2}%
{pd-masse2_err2}{-2}%
{pd-masse2_fmt}{$32 \pm 2$}%
{pd-masse3}{10.8}%
{pd-masse3_err1}{0.6}%
{pd-masse3_err2}{-0.6}%
{pd-masse3_fmt}{$10.8 \pm 0.6$}%
{pd-per1}{2.5081}%
{pd-per1_err1}{0.0002}%
{pd-per1_err2}{-0.0002}%
{pd-per1_fmt}{$2.5081 \pm 0.0002$}%
{pd-per2}{7.9222}%
{pd-per2_err1}{0.0001}%
{pd-per2_err2}{-0.0001}%
{pd-per2_fmt}{$7.9222 \pm 0.0001$}%
{pd-per3}{11.8993}%
{pd-per3_err1}{0.0008}%
{pd-per3_err2}{-0.0008}%
{pd-per3_fmt}{$11.8993 \pm 0.0008$}%
{pd-tc1}{2021.0726}%
{pd-tc1_err1}{0.0018}%
{pd-tc1_err2}{-0.0018}%
{pd-tc1_fmt}{$2021.0726 \pm 0.0018$}%
{pd-tc2}{2027.9023}%
{pd-tc2_err1}{0.0002}%
{pd-tc2_err2}{-0.0002}%
{pd-tc2_fmt}{$2027.9023 \pm 0.0002$}%
{pd-tc3}{2020.0007}%
{pd-tc3_err1}{0.0004}%
{pd-tc3_err2}{-0.0004}%
{pd-tc3_fmt}{$2020.0007 \pm 0.0004$}%
{pd-secosw1}{0.00}%
{pd-secosw1_err1}{0.00}%
{pd-secosw1_err2}{-0.00}%
{pd-secosw1_fmt}{$0.00 \pm 0.00$}%
{pd-secosw2}{0.02}%
{pd-secosw2_err1}{0.06}%
{pd-secosw2_err2}{-0.06}%
{pd-secosw2_fmt}{$0.02 \pm 0.06$}%
{pd-secosw3}{0.04}%
{pd-secosw3_err1}{0.04}%
{pd-secosw3_err2}{-0.04}%
{pd-secosw3_fmt}{$0.04 \pm 0.04$}%
{pd-sesinw1}{0.00}%
{pd-sesinw1_err1}{0.00}%
{pd-sesinw1_err2}{-0.00}%
{pd-sesinw1_fmt}{$0.00 \pm 0.00$}%
{pd-sesinw2}{-0.44}%
{pd-sesinw2_err1}{0.04}%
{pd-sesinw2_err2}{-0.04}%
{pd-sesinw2_fmt}{$-0.44 \pm 0.04$}%
{pd-sesinw3}{-0.46}%
{pd-sesinw3_err1}{0.03}%
{pd-sesinw3_err2}{-0.03}%
{pd-sesinw3_fmt}{$-0.46 \pm 0.03$}%
{pd-inc1}{90.8}%
{pd-inc1_err1}{0.7}%
{pd-inc1_err2}{-0.7}%
{pd-inc1_fmt}{$90.8 \pm 0.7$}%
{pd-inc2}{91.5}%
{pd-inc2_err1}{0.1}%
{pd-inc2_err2}{-0.1}%
{pd-inc2_fmt}{$91.5 \pm 0.1$}%
{pd-inc3}{91.1}%
{pd-inc3_err1}{0.1}%
{pd-inc3_err2}{-0.1}%
{pd-inc3_fmt}{$91.1 \pm 0.1$}%
{pd-Omega1}{0.0}%
{pd-Omega1_err1}{0.0}%
{pd-Omega1_err2}{-0.0}%
{pd-Omega1_fmt}{$0.0 \pm 0.0$}%
{pd-Omega2}{0.0}%
{pd-Omega2_err1}{0.0}%
{pd-Omega2_err2}{-0.0}%
{pd-Omega2_fmt}{$0.0 \pm 0.0$}%
{pd-Omega3}{-7.4}%
{pd-Omega3_err1}{0.8}%
{pd-Omega3_err2}{-0.8}%
{pd-Omega3_fmt}{$-7.4 \pm 0.8$}%
{pd-omegadeg1}{0}%
{pd-omegadeg1_err1}{0}%
{pd-omegadeg1_err2}{-0}%
{pd-omegadeg1_fmt}{$0 \pm 0$}%
{pd-omegadeg2}{-88}%
{pd-omegadeg2_err1}{8}%
{pd-omegadeg2_err2}{-8}%
{pd-omegadeg2_fmt}{$-88 \pm 8$}%
{pd-omegadeg3}{-85}%
{pd-omegadeg3_err1}{6}%
{pd-omegadeg3_err2}{-6}%
{pd-omegadeg3_fmt}{$-85 \pm 6$}%
{pd-omegadiffdeg}{2}%
{pd-omegadiffdeg_err1}{2}%
{pd-omegadiffdeg_err2}{-2}%
{pd-omegadiffdeg_fmt}{$2 \pm 2$}%
{pd-e1}{0.00}%
{pd-e1_err1}{0.00}%
{pd-e1_err2}{-0.00}%
{pd-e1_fmt}{$0.00 \pm 0.00$}%
{pd-e2}{0.20}%
{pd-e2_err1}{0.03}%
{pd-e2_err2}{-0.03}%
{pd-e2_fmt}{$0.20 \pm 0.03$}%
{pd-e3}{0.21}%
{pd-e3_err1}{0.03}%
{pd-e3_err2}{-0.03}%
{pd-e3_fmt}{$0.21 \pm 0.03$}%
{pd-ror1}{0.0124}%
{pd-ror1_err1}{0.0004}%
{pd-ror1_err2}{-0.0004}%
{pd-ror1_fmt}{$0.0124 \pm 0.0004$}%
{pd-ror2}{0.0777}%
{pd-ror2_err1}{0.0006}%
{pd-ror2_err2}{-0.0006}%
{pd-ror2_fmt}{$0.0777 \pm 0.0006$}%
{pd-ror3}{0.0458}%
{pd-ror3_err1}{0.0004}%
{pd-ror3_err2}{-0.0004}%
{pd-ror3_fmt}{$0.0458 \pm 0.0004$}%
{pd-c1}{0.4}%
{pd-c1_err1}{0.1}%
{pd-c1_err2}{-0.1}%
{pd-c1_fmt}{$0.4 \pm 0.1$}%
{pd-c2}{0.3}%
{pd-c2_err1}{0.2}%
{pd-c2_err2}{-0.2}%
{pd-c2_fmt}{$0.3 \pm 0.2$}%
{pd-prad1}{1.11}%
{pd-prad1_err1}{0.05}%
{pd-prad1_err2}{-0.05}%
{pd-prad1_fmt}{$1.11 \pm 0.05$}%
{pd-prad2}{7.0}%
{pd-prad2_err1}{0.2}%
{pd-prad2_err2}{-0.2}%
{pd-prad2_fmt}{$7.0 \pm 0.2$}%
{pd-prad3}{4.1}%
{pd-prad3_err1}{0.2}%
{pd-prad3_err2}{-0.2}%
{pd-prad3_fmt}{$4.1 \pm 0.2$}%
{pd-masseprec1}{5.07}%
{pd-masseprec1_err1}{2.40}%
{pd-masseprec1_err2}{-2.40}%
{pd-masseprec1_fmt}{$5.07 \pm 2.40$}%
{pd-masseprec2}{32.4}%
{pd-masseprec2_err1}{1.7}%
{pd-masseprec2_err2}{-1.7}%
{pd-masseprec2_fmt}{$32.4 \pm 1.7$}%
{pd-masseprec3}{10.80}%
{pd-masseprec3_err1}{0.57}%
{pd-masseprec3_err2}{-0.57}%
{pd-masseprec3_fmt}{$10.80 \pm 0.57$}%
{lopez-fenv2}{44}%
{lopez-fenv2_err1}{3}%
{lopez-fenv2_err2}{-3}%
{lopez-fenv2_fmt}{$44 \pm 3$}%
{lopez-mcore2}{18}%
{lopez-mcore2_err1}{1}%
{lopez-mcore2_err2}{-1}%
{lopez-mcore2_fmt}{$18 \pm 1$}%
{lopez-fenv3}{14}%
{lopez-fenv3_err1}{1}%
{lopez-fenv3_err2}{-1}%
{lopez-fenv3_fmt}{$14 \pm 1$}%
{lopez-mcore3}{9.4}%
{lopez-mcore3_err1}{0.5}%
{lopez-mcore3_err2}{-0.5}%
{lopez-mcore3_fmt}{$9.4 \pm 0.5$}%
{lopez-teq1}{1247}%
{lopez-teq1_err1}{38}%
{lopez-teq1_err2}{-38}%
{lopez-teq1_fmt}{$1247 \pm 38$}%
{lopez-teq2}{850}%
{lopez-teq2_err1}{26}%
{lopez-teq2_err2}{-26}%
{lopez-teq2_fmt}{$850 \pm 26$}%
{lopez-teq3}{742}%
{lopez-teq3_err1}{23}%
{lopez-teq3_err2}{-23}%
{lopez-teq3_fmt}{$742 \pm 23$}%
}[XX]%
}

%% file: tab_times.tex
K2-19b & 30 & FLWO & 2218.0041 & 0.0022 & B \\
K2-19b & 34 & TRAPPIST & 2249.6955 & 0.0014 & B \\
K2-19b & 41 & MuSCAT & 2305.1505 & 0.0014 & B \\
K2-19b & 133 & Spitzer & 3033.8604 & 0.0009 & A \\
K2-19c & 87 & Spitzer & 3019.4774 & 0.0074 & A \\
K2-19b & 141 & LCO & 3097.2502 & 0.0024 & A \\
K2-19b & 150 & Spitzer & 3168.5368 & 0.0014 & A \\
K2-19c & 102 & Spitzer & 3197.8645 & 0.0059 & A \\

%% file: tab_rv-stub.tex
2225.996346 & -8.92 & 2.69 & 0.358 \\
2229.058283 & -14.53 & 2.84 & 0.328 \\
2346.849965 & -11.35 & 1.98 & 0.181 \\
2366.792920 & -0.21 & 2.04 & 0.247 \\
2367.829151 & -9.39 & 3.37 & 0.221 \\
2368.814357 & -14.07 & 2.20 & 0.182 \\
2370.809676 & -18.62 & 2.13 & 0.221 \\
2374.805352 & 6.22 & 3.15 & 0.195 \\
2375.803685 & 2.94 & 2.18 & 0.249 \\
2376.797458 & -14.34 & 2.12 & 0.269 \\

%% file: tab_transit-times-predict-stub1.tex
b & 0 & 2014-06-04 & 1980.3840 & 0.0002 \\
c & 0 & 2014-06-08 & 1984.2722 & 0.0008 \\
b & 1 & 2014-06-12 & 1988.3041 & 0.0002 \\
c & 1 & 2014-06-20 & 1996.1834 & 0.0006 \\
b & 2 & 2014-06-20 & 1996.2220 & 0.0002 \\

%% file: tab_transit-times-predict-stub2.tex
c & 476 & 2029-12-11 & 7648.8365 & 0.1814 \\
b & 716 & 2029-12-14 & 7651.5243 & 0.0468 \\
b & 717 & 2029-12-21 & 7659.4466 & 0.0446 \\
c & 477 & 2029-12-23 & 7660.7298 & 0.1710 \\
b & 718 & 2029-12-29 & 7667.3662 & 0.0434 \\